\documentclass{aa}
\usepackage{natbib}
\usepackage{txfonts}
\usepackage{hyperref}
\usepackage{graphicx}
\usepackage{subfig}
\graphicspath{{figures/}}

\newcommand{\grad}{\vec\nabla}

\begin{document}
 
\title{Asymptotic theory of gravity modes in rotating stars}
\subtitle{II.~Impact of general differential rotation}
\titlerunning{Asymptotic theory of gravity modes in rotating stars. II.~Impact of general differential rotation}

\author{V. Prat\inst{1,2} \and S. Mathis\inst{1,2} \and K. Augustson\inst{1,2} \and F. Ligni\`eres \inst{3,4} \and J. Ballot \inst{3,4} \and L. Alvan\inst{1,2} \and A. S. Brun\inst{1,2}}

\institute{
    IRFU, CEA, Universit\'e Paris-Saclay, F-91191 Gif-sur-Yvette, France    
    \and
    Universit\'e Paris Diderot, AIM, Sorbonne Paris Cit\'e, CEA, CNRS, F-91191 Gif-sur-Yvette, France
    \and
    Universit\'e de Toulouse; UPS-OMP; IRAP; Toulouse, France
    \and
    CNRS; IRAP; 14 avenue \'Edouard Belin; F-31400 Toulouse, France
}

\date{}

\abstract
{Differential rotation has a strong influence on stellar internal dynamics and evolution, notably by triggering hydrodynamical instabilities, by interacting with the magnetic field, and more generally by inducing transport of angular momentum and chemical elements.
Moreover, it modifies the way waves propagate in stellar interiors and thus the frequency spectrum of these waves, the regions they probe, and the transport they generate.}
{We investigate the impact of a general differential rotation (both in radius and latitude) on the propagation of axisymmetric gravito-inertial waves.}
{We use a small-wavelength approximation to obtain a local dispersion relation for these waves.
We then describe the propagation of waves thanks to a ray model that follows a Hamiltonian formalism.
Finally, we numerically probe the properties of these gravito-inertial rays for different regimes of radial and latitudinal differential rotation.}
{We derive a local dispersion relation that includes the effect of a general differential rotation.
Subsequently, considering a polytropic stellar model, we observe that differential rotation allows for a large variety of resonant cavities that can be probed by gravito-inertial waves.
We identify that for some regimes of frequency and differential rotation, the properties of gravito-inertial rays are similar to those found in the uniformly rotating case.
Furthermore, we also find new regimes specific to differential rotation, where the dynamics of rays is chaotic.}
{As a consequence, we expect modes to follow the same trend.
Some parts of oscillation spectra corresponding to regimes similar to those of the uniformly rotating case would exhibit regular patterns, while parts corresponding to the new regimes would be mostly constituted of chaotic modes with a spectrum rather characterised by a generic statistical distribution.}


\maketitle

\section{Introduction}
\label{sec:intro}

Differential rotation plays a key role in stellar dynamics, magnetism, and evolution.
It triggers instabilities (such as Rayleigh-Taylor, Goldreich-Schubert-Friecke, or shear instabilities) and interacts with large-scale flows (such as meridional circulation) that induce transport of chemicals and angular momentum in stellar interiors \citep[e.g.][]{Zahn92, MaederZahn, MathisZahn}.
This transport greatly modifies the structural, rotational, chemical and magnetic evolution of stars \citep[e.g.][and references therein]{Maeder, Mathis13, Rieutord06}.
It is therefore crucial to constrain the amount of differential rotation present in stars to better understand these transport processes.
Further, differential rotation can drastically impact the propagation and the frequency spectrum of gravity waves, especially in the case of strong gradients of angular velocity \citep{Ando, LeeSaio93, Mathis09, Mirouh, Guenel}.
This is of great importance for stars that exhibit gravity waves, since these waves allow for probing stellar interiors and rotation thanks to seismic diagnoses based on the asymptotic properties of the waves \citep[e.g.][]{Bouabid, PratMLBC}.
Moreover, these waves are able to transport angular momentum.
In particular, they provide a mechanism to explain the weak differential rotation revealed by the CoRoT and Kepler space missions in solar-type \citep{Schatzman, Zahn97, TalonCharbonnel05}, evolved \citep{TalonCharbonnel08, Fuller, Pincon} and massive stars \citep{Lee, Rogers13, Rogers15}.

Seismology of slow rotators, in which rotation can be considered as a perturbation of the non-rotating case, provides us with many constraints on their internal structure and rotation.
Within this approximation, the main effect of rotation on oscillations is to lift the degeneracy between modes of the same radial order and degree but different azimuthal orders \citep{Saio81}.
These rotational splittings depend directly on the internal rotation profile of the stars \citep[and references therein]{Aerts10}.

In the case of the Sun, splittings of solar \emph{p} modes have highlighted a latitudinal differential rotation in the convective envelope and a relatively flat rotation profile in the radiative zone down to $0.2R_\odot$ \citep{Thompson, Couvidat}. 
Splittings of \emph{g}-mode candidates also suggested that the solar core rotates around four times as fast as the bulk of the radiative zone \citep{Garcia07}.
This has recently been confirmed by \citet{Fossat} using frequency modulations of \emph{p} modes by low-frequency \emph{g} modes.

For distant stars, which we cannot spatially resolve, progress has also been made in this direction.
Many subgiant and red giant stars exhibit rotational splittings of mixed modes that allow us to estimate the contrast in rotation between their core and their surface \citep{Beck, Mosser, Deheuvels12, Deheuvels14, Deheuvels15, Triana17}.
It is found that the core of such stars typically rotates between 2 and 20 times as fast as their envelope.
Constraints on differential rotation inside solar-type stars can be put by combining information from rotational splittings with estimates of the surface rotation rate obtained using other methods \citep{Benomar, Nielsen}.
In some more massive stars, rotational splittings of \emph{g} modes provide constraints on their internal rotation \citep{Triana15, Murphy}.
When \emph{p}-mode splittings are also observed, the contrast in rotation between the core and the envelope can be estimated \citep{Kurtz, Saio15}.
Some of these stars exhibit a core that rotates slower than the envelope, down to a rotation ratio of 30\%.

For fast rotators, such as $\gamma$ Doradus, $\delta$ Scuti, SPB, $\beta$ Cephei or Be stars, however, perturbative methods fail \citep{Reese06, Ballot10, Ballot13} and a more complete treatment of rotation is needed.
The eigenvalue problem is fully two-dimensional (2D), in the sense that it cannot be separated into one-dimensional problems as in the non-rotating case \citep{Rieutord09}.
In general, the problem is not solvable analytically and is computationally more expensive than separable problems.
A first attempt to propose new diagnoses for \emph{g} modes based on 2D computations of modes has been made by \citet{Ouazzani17}.

To simplify the problem of computing eigenmodes for moderate rotators, the so-called traditional approximation \citep{Eckart} can be used.
It consists in neglecting the horizontal component of the rotation vector, which makes the problem separable again if the star is assumed to rotate uniformly and if the centrifugal deformation is not taken into account.
Within this approximation, the effect of the rotation on low-frequency \emph{g} modes is greatly simplified \citep{LeeSaio97, Townsend03, Bouabid} and the approximation has been used to interpret seismic data of $\gamma$ Doradus \citep{VanReeth} and SPB stars \citep{Papics17}.

The effect of differential rotation on the waves and on the transport they generate has been studied by various authors \citep{Ando, LeeSaio93, DzhalilovStaude, Mathis09} but most of them focus on a given type of differential rotation profile (e.g. cylindrical or radial), or use simplifying assumptions that are not completely justified, such as the traditional approximation.

To progress in the understanding of the effect of a general differential rotation on stellar oscillations, it is also possible to build asymptotic theories, which make approximations to gain physical insight.
Studies of gravito-inertial waves based on characteristics have been done for uniform \citep{FriedlanderSiegmann, DintransRieutord} and differential \citet{Mirouh} rotation.
In the case of purely inertial waves (without stratification), radial and cylindrical differential rotation profiles \citep{BaruteauRieutord} and conical ones \citep{Guenel} have been considered.

Contrary to the method of characteristics, the small-wavelength Jeffrey-Wentzel-Kramers-Brillouin (JWKB) asymptotic theory can take into account the compressibility effects that produce the back-refraction of the waves approaching the stellar surface.
This approach has been followed by \citep{LG09} to study acoustic modes in rapidly rotating stars.
Using semi-classical quantisation concepts to link acoustic rays to pressure modes, they predicted spectral patterns that have been successfully confronted to bi-dimensional numerical computations of modes \citep[see also][]{Pasek11, Pasek12}.

More recently, a JWKB asymptotic theory for \emph{g} modes in a uniformly rotating star was built \citep[hereafter referred to as Paper~I]{PratLB}.
For the first time, this theory took into account the full effect of rotation (both Coriolis and centrifugal accelerations) and a realistic back-refraction of gravito-inertial waves near the stellar surface.
The main prediction of this paper is that modes can be classified as either (i) regular modes, which are similar to modes found in the non-rotating case; (ii) island modes, where the energy is localised around a so-called periodic orbit; or (iii) chaotic modes, which are expected to be characterised by an irregular spatial pattern and a generic statistical distribution of the frequency spacings between nearest modes.
Using results of Paper I, \citet{PratMLBC} derived theoretical period spacings in the low-frequency regime, where regular modes are expected to dominate.

The aim of the current paper is to investigate the impact of general differential rotation on axisymmetric gravito-inertial waves.
First, we derive an eikonal equation (i.e. a local dispersion relation) for axisymmetric gravito-inertial waves in a differentially rotating star (Sect.~\ref{sec:eikonal}).
Second, we describe the ray model associated with the eikonal equation and the tools used to numerically investigate the nature of the ray dynamics (Sect.~\ref{sec:raymodel}).
Third, we explore the ray dynamics for various types of differential rotation profiles in Sect.~\ref{sec:dynamics}.
Finally, we conclude in Sect.~\ref{sec:discussion}.

\section{Eikonal equation for gravito-inertial waves}
\label{sec:eikonal}

\subsection{Baroclinic, rotating equilibrium model}
\label{sec:poly}

We consider here a general background model of star that takes baroclinicity into account.
The equilibrium equations are
\begin{align}
    -\rho_0 s\Omega^2\vec e_s   &= -\grad P_0 - \rho_0\grad\psi_0,    \label{eq:back_mom}   \\
    \nabla^2\psi_0              &= 4\pi G\rho_0,                    \label{eq:poisson} 
\end{align}
where $\rho_0$, $P_0$, and $\psi_0$ are the equilibrium density, pressure and gravitational potential, respectively; $s$ is the cylindrical radial coordinate, $\vec e_s$ the unit vector associated with it, $\Omega$ is the rotation rate, and $G$ the gravitational constant.
Equation of state and an entropy equation have to be added to these equations to close the system.
The effects of an induced meridional circulation are neglected.
In practice, Eqs.~\eqref{eq:back_mom} and~\eqref{eq:poisson} are sufficient to specify the background terms that are involved in the dispersion relation of waves.

Equation~\eqref{eq:back_mom} can be written in the form
\begin{equation}
    \label{eq:geff}
    0=-\grad P_0+\rho_0\vec g_0,
\end{equation}
where $\vec g_0$ is the effective gravity defined by
\begin{equation}
    \label{eq:defg}
    \vec g_0=-\grad\psi_0+s\Omega^2\vec e_s.
\end{equation}
In the barotropic case, pressure is a function of density only, so that $\vec g_0$ can be defined as a gradient, as done in Paper I; in the baroclinic case, $\vec g_0$ is no longer a gradient.

\subsection{Perturbation equations}
\label{sec:perturb}

We use here two major approximations: the adiabatic approximation, which neglects dissipative processes so that the evolution of the fluid is isentropic, and the Cowling approximation, which neglects the variations of the gravitational potential induced by waves \citep{Cowling}.
The equations for small perturbations around the equilibrium state in the inertial frame are
\begin{align}
    (\partial_t+\Omega\partial_\varphi)\rho+\grad\cdot(\rho_0\vec v)                                    &=  0,  \label{eq:cont}\\
    (\partial_t+\Omega\partial_\varphi)v_i+(\vec f\wedge\vec v)_i+(\vec v\cdot\vec Q)\delta_{\varphi i} &=  -\frac{\nabla_i P}{\rho_0} + \frac{\rho}{\rho_0}g_{0i},   \label{eq:mom_pert}\\
    {c_{\rm s}}^2\left[(\partial_t+\Omega\partial_\varphi)\rho+\vec v\cdot\grad\rho_0\right]            &=  (\partial_t+\Omega\partial_\varphi)P+\vec v\cdot\grad P_0,  \label{eq:iso}
\end{align}
where
\begin{align}
    \vec f  &=  2\Omega\vec e_z,    \\
    \vec Q  &=  s\grad\Omega,
\end{align}
$\rho$, $\vec v$, and $P$ are density, velocity, and pressure fluctuations, respectively; $c_{\rm s}$ is the sound speed, and $\vec e_z$ and $\vec e_\varphi$ are the unit vectors aligned with the rotation axis and in the azimuthal direction (associated with the azimuthal angle $\varphi$), respectively.
As shown in Appendix~\ref{sec:deriv}, it is possible to obtain a single equation for pressure fluctuations only of the form
\begin{equation}
    \label{eq:wave_gen}
    \vec{\mathcal{A}}:\grad\grad\hat P+\vec B\cdot\grad\hat P+C\hat P =0,
\end{equation}
where $\vec{\mathcal A}$ is a rank-2 tensor; the double gradient of a scalar function $a$ is defined by $(\vec\nabla\vec\nabla a)_{ij}=\partial_i\partial_ja$ in Cartesian coordinates; the symbol $:$ denotes the double contraction of two tensors ($\vec{\mathcal{X}}:\vec{\mathcal{Y}}=\sum_{ij}\mathcal{X}^{ij}\mathcal{Y}_{ji}$); and $\hat P$ is the complex amplitude of pressure fluctuations, such that
\begin{equation}
    P(\vec x,t)=\Re\left[\hat P(r, \theta)e^{i(m\varphi-\omega t)}\right],
\end{equation}
where $m$ is the azimuthal order of the waves and $\omega$ their angular frequency.
Although the problem is initially three-dimensional (3D), for a given $m$, it becomes 2D due to the quantisation in the azimuthal direction.

\subsection{Short-wavelength approximation}
\label{sec:wkb}

After some calculations explained in Appendix~\ref{sec:surface}, the wave equation~\eqref{eq:wave_gen} is transformed into a local dispersion relation, also called an eikonal equation.
This is done by using the JWKB approximation, which consists in searching for wave-like solutions of the form
\begin{equation}
    \hat P(\vec x) = \Re\left[A(\vec x)e^{i\Phi(\vec x)}\right],
\end{equation}
where the wavelength associated with the wavevector $\vec k=\grad\Phi$ ($\Phi$ is the phase) is much smaller than the typical length of variations of the amplitude $A$ associated with the variations of the background model.
As discussed in Paper~I, this would normally be equivalent to retaining only second-order derivatives, with $(\vec\nabla\vec\nabla\hat P)_{ij}\simeq -\hat Pk_ik_j$.
However, to ensure that waves are back-refracted near the stellar surface (which is needed to study modes) we also retain a zeroth-order term that becomes large near the surface.
This means that the JWKB approximation is valid only in an intermediate regime of wavelength, where the latter has to be larger than the inverse of the density scaleheight.
Moreover, since we focus here on the asymptotic regime of low-frequency gravity waves, we can neglect as a first step the acoustic part of the wave equation.

\begin{figure}
    \resizebox{\hsize}{!}{\includegraphics{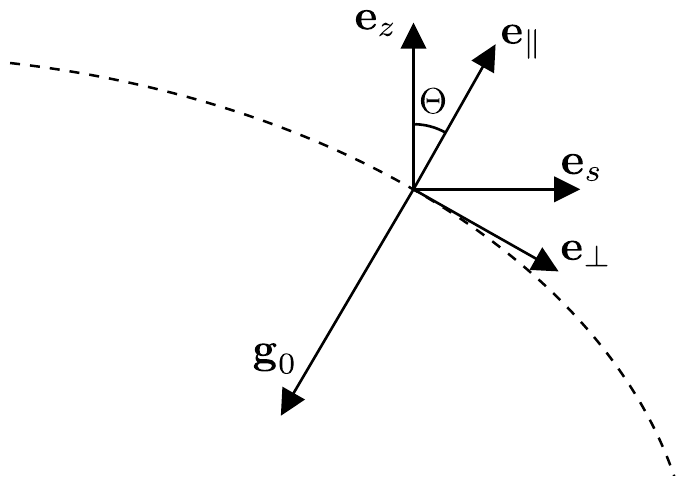}}
    \caption{Coordinate systems }
    \label{fig:coords}
\end{figure}

There is \emph{a priori} no reason to assume that $m$ is large enough so that terms in $m^2$ or $m\vec k$ should also be retained in the eikonal equation.
The latter then reads
\begin{equation}
    \begin{aligned}
        (k^2+{k_{\rm c}}^2){\hat\omega}^2 &= f(f+Q_s){k_z}^2+{N_0}^2{k_\perp}^2-fQ_z(k_sk_z+k_\parallel k_\perp)    \\
                                    &\quad +f\cos\Theta(f\cos\Theta+Q_\perp){k_{\rm c}}^2,
    \end{aligned}
\end{equation}
where $\hat\omega=\omega-m\Omega$ is the Doppler-shifted angular frequency, $\parallel$ and $\perp$ refer to components along $\vec e_\parallel$ and $\vec e_\perp$, unit vectors parallel and orthogonal to the effective gravity, and $\Theta$ is the angle between the rotation axis and $\vec e_\parallel$, as illustrated in Fig.~\ref{fig:coords}; and $k_{\rm c}$, defined in Eq.~\eqref{eq:kc_fin}, is a term accounting for the back-refraction of waves near the stellar surface.
The only dependence on $m$ is included in the Doppler-shifted frequency.

It is possible, however, to obtain a more complete eikonal equation by considering that the aforementioned terms in $m^2$ or $m\vec k$ should be retained as well because they come from second-order derivatives (including those involving $\varphi$).
Subsequently, the eikonal equation becomes
\begin{equation}
    \label{eq:eik_full}
    \begin{aligned}
    &i\hat\omega^3(k^2+{k_{\rm c}}^2) + \hat\omega^2k_\varphi\vec Q\cdot\vec k \\
    &\quad
        \begin{aligned}
            -i\hat\omega&\left[f(f+Q_s){k_z}^2+{N_0}^2({k_\perp}^2+{k_\varphi}^2)\right. \\
                        &\quad\left.-fQ_z(k_sk_z+k_\parallel k_\perp) + f\cos\Theta(f\cos\Theta+Q_\perp){k_{\rm c}}^2\right]
        \end{aligned}   \\
        &\quad +k_\varphi\left[(fQ_zQ_\parallel-{N_0}^2Q_\perp)k_\perp- f^2Q_zk_z\right]=0,
    \end{aligned}
\end{equation}
where $k_\varphi=m/(r\sin\theta)$. 
In the general case, this equation can be solved analytically for $\hat\omega$, but the solutions are non-trivial functions of structure, rotation, and $\vec k$.

When there is no differential rotation, we recover the eikonal equation obtained in Paper~I:
\begin{equation}
    \hat\omega^2=\frac{f^2{k_z}^2+{N_0}^2({k_\perp}^2+{k_\varphi}^2)+f^2\cos^2\Theta{k_{\rm c}}^2}{k^2+{k_{\rm c}}^2}.
\end{equation}
When only axisymmetric waves are considered ($k_\varphi=0$, and $\hat\omega=\omega$), Eq.~\eqref{eq:eik_full} reduces to
\begin{equation}
    \label{eq:eik_axi}
    \begin{aligned}
        (k^2+{k_{\rm c}}^2)\omega^2 &= f(f+Q_s){k_z}^2+{N_0}^2{k_\perp}^2-fQ_z(k_sk_z+k_\parallel k_\perp)    \\
                                    &\quad +f\cos\Theta(f\cos\Theta+Q_\perp){k_{\rm c}}^2.
    \end{aligned}
\end{equation}
When centrifugal deformation (which also includes effects of baroclinicity) is neglected, the eikonal equation in spherical coordinates reads
\begin{equation}
    \label{eq:eik_sph}
    \begin{aligned}
        (k^2+{k_{\rm c}}^2)\omega^2 &= ({k_r}^2+{k_{\rm c}}^2)f\cos\theta(f\cos\theta+Q_\theta)   \\
                                    &\quad+{k_\theta}^2\left[{N_0}^2+f\sin\theta(f\sin\theta+Q_r)\right]    \\
                                    &\quad-k_rk_\theta\left[f\cos\theta(f\sin\theta+Q_r)\right.\\
                                    &\quad\quad\left.+f\sin\theta(f\cos\theta+Q_\theta)\right].
    \end{aligned}
\end{equation}
From now on, we focus our study on axisymmetric waves that are described by the eikonal equation~\eqref{eq:eik_axi}.

\subsection{Domains of propagation}
\label{sec:domain}

The eikonal equation~\eqref{eq:eik_axi} can be seen as a quadratic equation in $k_\parallel$ or $k_\perp$.
It is thus possible to derive a condition for them to be real:
\begin{equation}
    \Gamma{k_\perp}^2-\left[\omega^2-f\cos\Theta(f\cos\Theta+Q_\perp)\right]^2{k_{\rm c}}^2 \geq 0,
\end{equation}
where
\begin{equation}
    \label{eq:gamma}
    \begin{aligned}
        \Gamma &= f\cos\Theta\left[fQ_z(f\sin\Theta+Q_\parallel)-{N_0}^2(f\cos\Theta+Q_\perp)\right]    \\
        &\quad\quad +\omega^2\left[{N_0}^2 + f(f+Q_s)\right]-\omega^4.
    \end{aligned}
\end{equation}
In the bulk of the star, $k_{\rm c}$ is negligible, so the propagation condition reduces to $\Gamma\geq0$.
In the following, we focus on the region where this is the case.
$\Gamma$ can be seen as a quadratic function of $\omega^2$ that always have two real solutions, ${\omega_-}^2$ and ${\omega_+}^2$, as proved in Appendix~\ref{sec:gamma}.
Thus, the propagation condition is equivalent to
\begin{equation}
    \label{eq:cond_root}
    {\omega_-}^2 \leq \omega^2 \leq {\omega_+}^2.
\end{equation}

In the particular case of a purely radial differential rotation and when the background model is assumed to be spherically symmetric, one can write the propagation condition (based on the eikonal equation given in Eq.~\eqref{eq:eik_sph}, which neglects all effects of centrifugal deformation, including baroclinic effects) in the form
\begin{equation}
    \label{eq:critlat}
A(r)\cos^4\theta+B(r)\cos^2\theta+C(r)\geq0,
\end{equation}
with
\begin{align}
    A(r) &= -\Omega^2{\hat{Q}_r}^2,   \\
    B(r) &= \Omega^2{\hat Q_r}^2 - f^2{N_0}^2-\omega^2f\hat Q_r,    \\
    C(r) &= \omega^2\left[{N_0}^2+f(f+\hat Q_r) - \omega^2\right],
\end{align}
where $\hat Q_r=r\partial_r\Omega$.
Roots of Eq.~\eqref{eq:critlat} correspond to critical co-latitudes, which delimit regions where waves can propagate.
Thus, when a critical co-latitude $\theta_{\rm c}$ exists, it satisfies
\begin{equation}
    \theta_{\rm c} = \arccos\sqrt{\frac{-B\pm\sqrt{B^2-4AC}}{2A}},
\end{equation}
which depends in general on the radius.

In the case of general differential rotation but in regions where $N_0$ is much larger than $f$ and $Q$ (the norm of $\vec Q$), the propagation condition~\eqref{eq:cond_root} simplifies to
\begin{equation}
    \label{eq:prop_n0}
    f\cos\Theta(f\cos\Theta+Q_\perp) \leq \omega^2 \leq {N_0}^2+f\sin\Theta(f\sin\Theta+Q_\parallel).
\end{equation}
The left inequality means that critical co-latitudes $\Theta_{\rm c}$ may exist for which $\omega^2=f\cos\Theta_{\rm c}(f\cos\Theta_{\rm c}+Q_\perp)$.
In solid-body rotation, they exist if and only if $\omega<f$.
Here, since $f$ and $Q_\perp$ vary with space, the picture is much more complex, and many different situations may occur, with either none, one, or several critical surfaces. 
By definition, $Q_\perp$ vanishes along the rotation axis.
Further, if the rotation is symmetric with respect to the equator, $Q_\perp$ also vanishes at the equator.
This means that the lower part of the condition of propagation \eqref{eq:prop_n0} becomes $0\leq\omega$ at the equator (always propagation) and $f\leq\omega$ at the poles.
As a consequence, if $\omega<f(\theta=0)$, waves cannot propagate near the rotation axis, and there is at least one critical surface in latitude between $\theta=0$ and $\theta=\pi/2$.
When critical pseudo co-latitudes exist, they verify
\begin{equation}
    \cos\Theta_{\rm c}=\frac{-Q_\perp\pm\sqrt{{Q_\perp}^2+4\omega^2}}{2f},
\end{equation}
which is possible only when $\omega^2\leq f(f+Q_\perp)$ if $\Theta\leq\pi/2$, or $\omega^2\leq f(f-Q_\perp)$ if $\Theta\geq\pi/2$.
Thus, if $\omega>\max_\theta[f(f+|Q_\perp|)]$, there is no critical surface in latitude.

In contrast, in regions where $f$ is much larger than $Q$ and $N_0$, Eq.~\eqref{eq:cond_root} becomes
\begin{equation}
    {N_0}^2\cos^2\Theta-A \leq \omega^2 \leq f(f+Q_s)+{N_0}^2\sin^2\Theta +A,
\end{equation}
where $A=fQ_z\sin\Theta\cos\Theta+{Q_z}^2\cos^4\Theta$.
By definition, $\vec Q$ vanishes at the centre.
As $N_0$ also vanishes, the right-hand inequality means that the waves cannot propagate to the centre when $\omega>f$.

To summarise the above discussion on the propagation domains, we find that waves that are sub-inertial throughout ($\omega<f$) avoid a region around the rotation axis while waves that are super-inertial throughout ($\omega>f$) avoid the stellar centre.
This property is verified in differentially rotating stars as in uniformly rotating ones, but new features appear when waves are not  sub- or super-inertial throughout.
For instance, propagation domains with multiple avoided regions in latitude are only possible in differentially rotating stars.
Examples of these new types of domains are illustrated in Sect.~\ref{sec:dynamics}.

\citet{Mirouh} have also studied propagation domains of gravito-inertial waves.
However, they considered a Boussinesq fluid between two concentric rigid spheres for one particular radial differential rotation, which makes the comparison with the present case not very instructive.

\section{Ray model for axisymmetric waves}
\label{sec:raymodel}

\subsection{Hamiltonian formalism}
\label{sec:Hamilton}

The angular frequency of waves $\omega$ is constant.
Thus, the scalar function $\omega(\vec x, \vec k)$ must remain constant when waves propagate and be equal to $\omega(\vec x_0, \vec k_0)$, which is set by the initial condition $ (\vec x_0, \vec k_0)$.
This implies that the propagation of waves can be described with the Hamiltonian formalism.
We define a ray as a trajectory tangent to the group velocity $\vec v_{\rm g}=\grad_{\vec k}\omega$, where $\grad_{\vec k}$ is the gradient with respect to $\vec k$.
It reads
\begin{equation}
    \label{eq:dx}
    \frac{{\rm d}\vec x}{{\rm d}t} = \grad_{\vec k}\omega.
\end{equation}
This choice is motivated by the fact that the group velocity characterises the transport of energy by the waves, whereas the phase velocity $\vec v_{\rm p}=\omega\vec k/k^2$ is the velocity of the wave front.
The constancy of $\omega$ then requires the evolution of the wavevector along the ray path to be governed by
\begin{equation}
    \label{eq:dk}
    \frac{{\rm d}\vec k}{{\rm d}t} = -\grad_{\vec x} \omega,
\end{equation}
where $\grad_{\vec x}$ is the spatial gradient.
Equations~\eqref{eq:dx} and~\eqref{eq:dk} thus have a Hamiltonian form, where the Hamiltonian is $\omega$.

In Paper~I, it was shown that the ray dynamics can be written in spherical coordinates $(r,\theta)$ as
\begin{align}
    \frac{{\rm d}r}{{\rm d}t}           &= \frac{\partial\omega}{\partial k_r},   \label{eq:dyn_r}\\
    \frac{{\rm d}\theta}{{\rm d}t}      &= \frac1r\frac{\partial\omega}{\partial k_\theta},   \label{eq:dyn_theta}\\
    \frac{{\rm d}k_r}{{\rm d}t}         &= -\frac{\partial\omega}{\partial r} + \frac{k_\theta}{r}\frac{\partial\omega}{\partial k_\theta}, \\
    \frac{{\rm d}k_\theta}{{\rm d}t}    &= -\frac1r\frac{\partial\omega}{\partial\theta} - \frac{k_\theta}{r}\frac{\partial\omega}{\partial k_r}.   \label{eq:dyn_ktheta}
\end{align}
The presence of the last term in each of the last two equations comes from the fact that the basis used for the wavevector ($\vec k=k_r\vec e_r+k_\theta\vec e_\theta$) is different from the natural one associated with spherical coordinates ($\vec k=k_r\vec e_r+k_\theta^{\rm nat}\vec e_\theta/r$).
These equations are made explicit in Appendix~\ref{sec:spherical}.

\subsection{Phase-space visualisation: the Poincar\'{e} surface of section}
\label{sec:pss}

Rays can be studied as dynamical systems with two degrees of freedom, so the phase space is four-dimensional $(r,\theta,k_r,k_\theta)$.
At a given frequency, all trajectories stay on a 3D space because $\omega$ is constant in time.
To visualise the structure of the phase space, it is convenient to consider the intersection of the latter with a given hyper-surface, usually defined by fixing one phase-space coordinate.
This is called a Poincar\'e section or surface of section (PSS).
To have a complete view of the structure of the phase space, the intersecting hyper-surface must be chosen so that most trajectories intersect it several times.
For our problem, the equatorial plane ($\theta=\pi/2$) appears as a logical choice, because most of the ray trajectories cross it, including low-frequency waves, which are trapped near the equatorial plane.
It is important to retain points for the PSS only when the intersecting surface is crossed from a given side (in our case, from the region where $\theta<\pi/2$).
This ensures that two trajectories never cross on a PSS.

In the following, PSS are represented in the plane $(r, k_r)$.
For consistency with Paper I, the radius is normalised by the equatorial radius of the star $R_{\rm e}$, while the radial wavevector is normalised by the quantity $N_{0,{\rm max}}^{\rm e}/(\omega R_{\rm e})$, where $N_{0,{\rm max}}^{\rm e}$ is the value of the inner maximum of the radial profile of the Brunt-V\"ais\"al\"a frequency in the equatorial plane (see Paper I for more detail, in particular the simple analytical description of the PSS in the non-rotating case).

\subsection{Numerical method}
\label{sec:numerics}

To properly characterise the dynamics of gravito-inertial rays of a system with differential rotation and a non-trivial background state such as those considered here, it is necessary to follow a sufficient number of rays of varying frequency.
In particular, to identify the different regions of the phase space, invariant tori and chaotic regions, each of those rays needs to be well-sampled both in time and space.

The spatial accuracy of the ray dynamics is limited by the method employed to compute the background medium, that is the model of rapidly rotating star.
For the simulations carried out here, the background thermodynamic state is computed assuming a polytropic equation of state and uniform rotation, thus taking into account the rotationally modified gravitational potential.
This state is expressed as the coefficients of a truncated spectral expansion in radius as Chebyshev polynomials and in latitude as Legendre polynomials, thereby limiting the accuracy of its spatial reconstruction.
Changes in the values of the background state due to the spectral truncation error are very small in absence of discontinuity in the background state.

Generating a statistically significant sample of intersections with a given PSS requires a numerical integrator that is both stable and accurate enough to follow the ray for a sufficiently long time.
This integrator must also be robust in regions where the equations become stiff, such as near the coordinate origin, as well as near turning surfaces like the stellar surface.
Fortuitously, the ray dynamics equations employed here have a symplectic or Hamiltonian character, as described in Sect.~\ref{sec:Hamilton}.
Therefore, we may appeal to an extant class of implicit symplectic integrators that have the property that the simplectic structure is preserved under the discrete map of the numerical method.
This means that the volume of the phase space for each ray is also preserved.

One particular set of implicit symplectic integrators are the Gauss-Legendre-Runge-Kutta (GLRK) methods.
To make this explicit for ray dynamics, one can formulate the problem as a coupled set of ordinary differential equations ${\rm d}\vec y/{\rm d}t = \vec f(\vec y)$ with the initial condition $\vec y(0) = \vec y_0$, which also selects the frequency of the ray, where $\vec y$ is the tuple $(\vec x, \vec k)$ and $\vec f(\vec y) = (\grad_{\vec k} \omega, -\grad_{\vec x} \omega)$.
In a discretised form, the $s$-stage implicit GLRK method yields
\begin{equation}
  \vec y_{n+1} = \vec y_n + h\sum_{j=1}^s b_j \vec f(\vec\xi_j), \label{eqn:glrky}
\end{equation}
where $h$ is the fixed step size.
The $\vec\xi_j$ are given implicitly by
\begin{equation}
  \vec \xi_j - \vec y_n - h\sum_{i=1}^s a_{ji} \vec f(\vec \xi_i) = 0. \label{eqn:glrkxi}
\end{equation}

The computation of the coefficients $a_{ij}$ and $b_j$ follows from a set of algebraic equations that themselves are derived from the coefficients of the Taylor expansion of $\vec \xi_j$ and $\vec f$ \citep{butcher63}.
Even so, there are unconstrained parameters that allow for the construction of a variety of RK methods with various error bounds.
Here the $c_j=\sum_ia_{ji}$ are those free parameters and are chosen to be the zeros of the shifted Legendre polynomial of degree $s$.
This choice leads to the symplectic nature of the method in a non-trivial way \citep[e.g.][]{sanzserna88}.
We have implemented this method with a choice of stage between $s=1$ and 5, as the Butcher tableau for these methods are widely available, where the symmetry of the coefficients implies that these methods have a truncation error of $\mathcal{O}(h^{2s})$.

Since the $\vec \xi_j$ are implicit and $\vec f$ is a non-linear function, we have implemented a Newton-Raphson solver to compute them.
To do this, we construct a new vector function $\vec X(\vec y_n, \vec \xi_j)$ that concatenates Eqs.~\eqref{eqn:glrkxi} as
\begin{align}
  X_{s(j-1)+k}^q &= \xi_{jk}^q - y_{nk} - h\sum_{i=1}^s a_{ji} f_k(\vec \xi_i^q), \\
  J^q\left(\vec y_n, \vec \xi_j^q\right)\delta \vec X^{q} &= - \vec X^q, \\
  \vec X^{q+1} &= \vec X^{q} + \delta \vec X^{q}.
\end{align}
For clarity in identifying the component index of $\vec X$, we note that $s$ is the total number of stages, $j$ is the index of the current stage, and $k$ is the index of the component of the tuple $\vec y$.
Furthermore, the index $q$ denotes the current step of the iterative solver and $J^q$ is the Jacobian of the function $\vec X$ evaluated at $\vec y_n$ and $\vec\xi_j^q$.
To reduce the cost of its computation, the derivatives of $\vec X$ used to form the Jacobian are finite differences, rather than analytical derivatives.
The iterative implicit solver necessarily places greater restrictions on the step size $h$ that follow from the properties of $\vec f$ and from the requirement of uniqueness of the solutions $\vec \xi_j$.
Hence, these restrictions will also vary depending upon the initial conditions of each ray, which means that choosing an appropriate $h$ is a nontrivial exercise that often has to be done heuristically.
Moreover, enforcing global uniqueness typically requires $h$ to be chosen to be smaller than one would otherwise expect.

We have further implemented an adaptive time stepping method to partially circumvent the time step restrictions.
To do this, we solve a modified Hamiltonian problem following the method described in \citet{hairer97}, but one that retains the symplectic nature of the original equations.
The idea behind this is that one has a step size $h = \chi(\vec y)\Delta t$, which corresponds to a remapping of the time variable as $\tau = \chi(\vec y) t$.
Since the bulk of the time stepping issues occur near the coordinate origin, we have chosen $\chi(\vec y)=\chi(r) \propto 1 - e^{-\lambda r}$ with $r$ being the radius and $\lambda$ a constant.
A modified Hamiltonian is constructed that depends upon $\chi$ as $\mathcal{K} = \chi(r) (\mathcal{H} - \mathcal{H}_0)$, where $\mathcal{H} = \omega(\vec y)$ and $\mathcal{H}_0 = \omega(\vec y_0)$.
This implies that the symplecticity-preserving equations of motion become
\begin{align}
    \frac{{\rm d}\vec x}{{\rm dt}} &= \chi\left(r\right)\grad_{\vec k} \mathcal{H}, \\
    \frac{{\rm d}\vec k}{{\rm dt}} &= -\chi\left(r\right)\grad_{\vec x} \mathcal{H} - \frac{\partial\chi\left(r\right)}{\partial r}\left(\mathcal{H}-\mathcal{H}_0\right).
\end{align}
Therefore, the only notable change in the method to accommodate these time step changes is in the right-hand-side vector function $\vec f$ of the GLRK method.

This new method has been compared to the fourth-order explicit Runge-Kutta method used in Paper~I and is more accurate.
However, it comes at the expense of more function evaluations, and thus a higher computational cost.

\section{Gravito-inertial ray dynamics in differentially rotating stars}
\label{sec:dynamics}

For simplicity, the background stellar structure we used for our numerical computations is a centrifugally deformed polytropic (so barotropic) model of a uniformly rotating star, and the rotation profile is given by a prescribed analytical formula presented hereafter.
This is not fully self-consistent because (i) the centrifugal deformation is slightly incorrect, and (ii) the rotation profiles considered here would lead to a baroclinic structure.
This may for example affect the profile of the Brunt-Väisälä frequency.
However, this simplification allows us to easily investigate the various types of general differential rotation.

In the following we present the general law of differential rotation we use (Sect.~\ref{sec:rot}) and then study the ray dynamics, considering the two cases of radial (Sect.~\ref{sec:core}) and latitudinal (Sect.~\ref{sec:env}) differential rotation separately.

\subsection{Rotation profile}
\label{sec:rot}

For the computations shown in the present paper, we considered the following three-zones rotation profile:
\begin{equation}
    \label{eq:orig_rot}
    \Omega_0(r,\theta) = \frac{\Omega_{\rm C}e^{-\alpha_{\rm C}(r-r_{\rm C})}+\Omega_{\rm R}+[\Omega_{\rm R}+\Omega_{\rm D}\cos(2\theta)]e^{\alpha_{\rm E}(r-r_{\rm E})}}{e^{-\alpha_{\rm C}(r-r_{\rm C})}+1+e^{\alpha_{\rm E}(r-r_{\rm E})}}.
\end{equation}
The core is characterised by an almost uniform rotation rate $\Omega_{\rm C}$ up to a radius $r_{\rm C}$, and $\alpha_{\rm C}$ is the steepness of the transition with the bulk of the radiative zone.
The latter rotates at a uniform rate $\Omega_{\rm R}$.
The envelope starts at a radius $r_{\rm E}$ and has a mean rotation rate $\Omega_{\rm R}$ and a latitudinal differential rotation characterised by $\Omega_{\rm D}$.
The steepness of the transition between the envelope and the bulk of the radiative zone is given by $\alpha_{\rm E}$.
In solar-type stars, the envelope is convective, but here, since we use a polytropic model with a single polytropic index, the full star is radiative.
$\Omega_{\rm D}<0$ corresponds to a solar differential rotation, with the equator faster than the poles, whereas $\Omega_{\rm D}>0$ corresponds to an anti-solar differential rotation.
Both configurations are supported by 3D numerical simulations \citep[see][]{BrunToomre, Brown, Matt, Gastine, Brun}.

At the centre of the star, this profile may not be well defined.
First, if $\Omega_{\rm D}\not=0$, it gives different values of $\Omega_0$ for different values of $\theta$.
Second, the presence of differential rotation causes the radial gradient of rotation to be non-zero at the centre.
To eliminate these pathological features, we regularise the rotation profile, adding the following corrections to the original profile:
\begin{equation}
    \Omega(r,\theta) = \Omega_0(r,\theta) - \gamma e^{-\alpha_{\rm E}r^2/2+\delta r}\cos(2\theta) + \varepsilon\frac{e^{-\alpha_{\rm C}r}}{\alpha_{\rm C}},
\end{equation}
where
\begin{align}
    \gamma      &=  \Omega_{\rm D}\frac{e^{-\alpha_{\rm E}r_{\rm E}}}{e^{\alpha_{\rm C}r_{\rm C}}+1+e^{-\alpha_{\rm E}r_{\rm E}}},    \\
    \delta      &=  \frac{\alpha_{\rm E}+(\alpha_{\rm C}+\alpha_{\rm E})e^{\alpha_{\rm C}r_{\rm C}}}{e^{\alpha_{\rm C}r_{\rm C}}+1+e^{-\alpha_{\rm E}r_{\rm E}}}, \\
    \varepsilon &=  (\Omega_{\rm R} - \Omega_{\rm C})\frac{\alpha_{\rm C}e^{\alpha_{\rm C}r_{\rm C}} + (\alpha_{\rm C} + \alpha_{\rm E})e^{\alpha_{\rm C}r_{\rm C}-\alpha_{\rm E}r_{\rm E}}}{(e^{\alpha_{\rm C}r_{\rm C}}+1+e^{-\alpha_{\rm E}r_{\rm E}})^2}.
\end{align}

We need to consider the stability of the rotation profile with respect to hydrodynamical instabilities \citep[e.g.][]{Zahn83, Maeder}.
In particular, we made sure that the profiles used in the present paper are Rayleigh-Taylor stable, that is, the quantity
\begin{equation}
    {N_\Omega}^2=f(f+Q_s)
\end{equation}
is positive in the whole star.
We also verified that the rotation profile is never centrifugally unstable.
In a spherical model, this means that the quantity $r\sin^2\theta\Omega(\Omega+r\partial_r\Omega)$ must always be smaller than gravity.
At the equator, the radial differential rotation is negligible, and the previous stability criterion reduces to $\Omega<\Omega_{\rm K}$, where
\begin{equation}
    \Omega_{\rm K}=\sqrt{\frac{GM}{{R_{\rm eq}}^3}}
\end{equation}
is the critical angular velocity and $M$ is the stellar mass.

\subsection{Radial differential rotation near the core}
\label{sec:core}

In this section we investigate the influence of radial differential rotation near the core on the structure of the phase space.
This is motivated by a significant number of observations showing a contrast in the rotation rate between the core and the envelope of stars (see Sect.~\ref{sec:intro}).

We take $\Omega_{\rm D}=0$ and $\Omega_{\rm R}/\Omega_{\rm K}=0.38$.
The two cases (fast or slow core) are studied in Sect.~\ref{sec:cfast} and Sect.~\ref{sec:cslow}, respectively.

\subsubsection{Fast core}
\label{sec:cfast}

\begin{figure}
    \resizebox{\hsize}{!}{\includegraphics{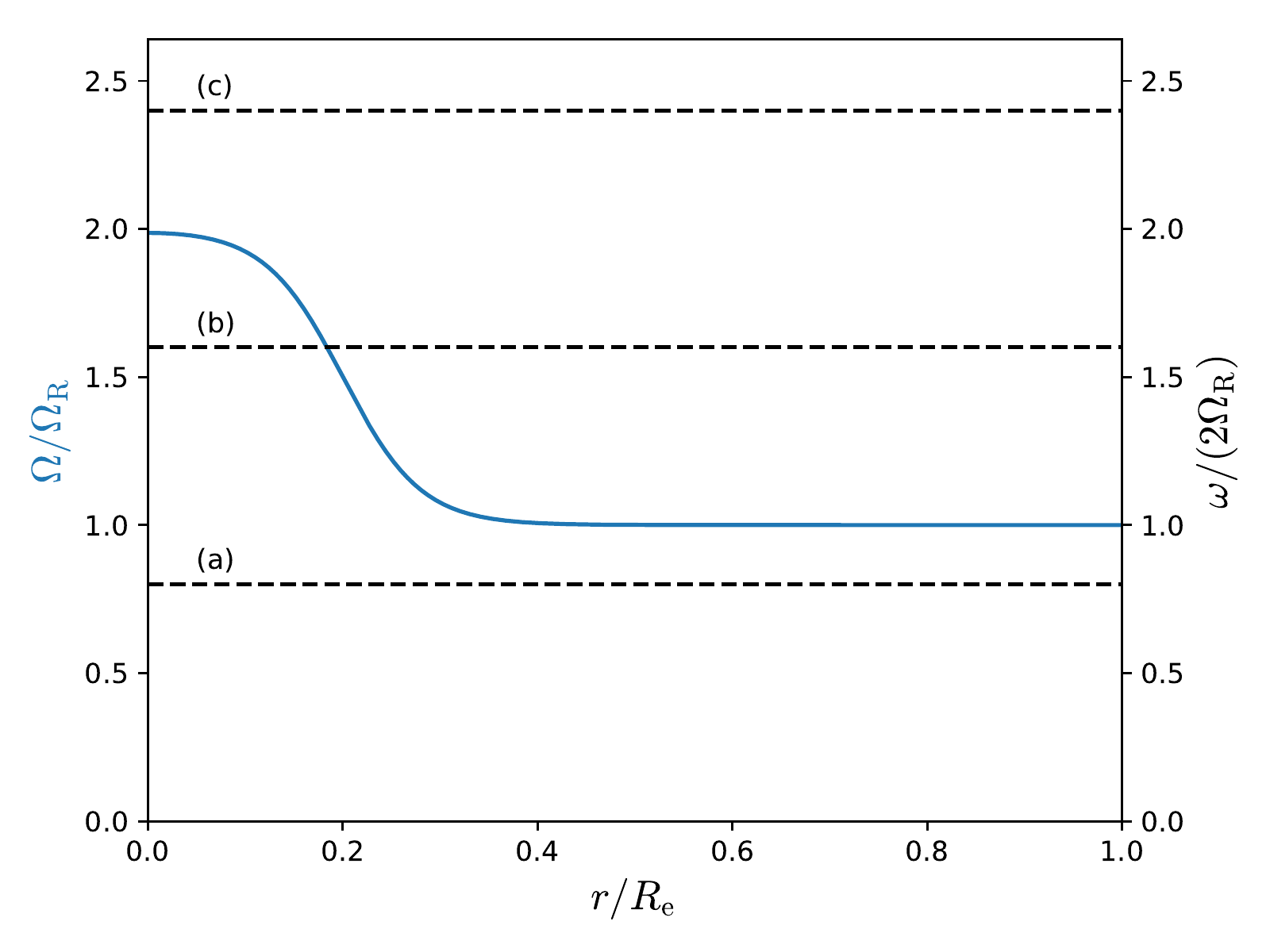}}
    \caption{Rotation profile with $\Omega_{\rm C}/\Omega_{\rm R}=2$, $r_{\rm C}=0.2$, and $\alpha_{\rm C}=25$ (blue solid line).
    The black dashed lines are the three frequencies for which we computed PSS: from bottom to top $\omega/(2\Omega_{\rm R})=0.8$, 1.6, and 2.4.
    They correspond to the sub-, trans-, and super-inertial regimes, respectively.}
    \label{fig:freq_oc2}
\end{figure}

First, we consider the case $\Omega_{\rm C}/\Omega_{\rm R}=2$.
We thus have $\Omega_{\rm R}<\Omega<\Omega_{\rm C}$, which defines three different frequency regimes.
To understand these regimes, we computed PSS at the three different frequencies shown in Fig.~\ref{fig:freq_oc2}.
\begin{figure*}
    \resizebox{\hsize}{!}{\subfloat{\label{fig:2_sub}\includegraphics{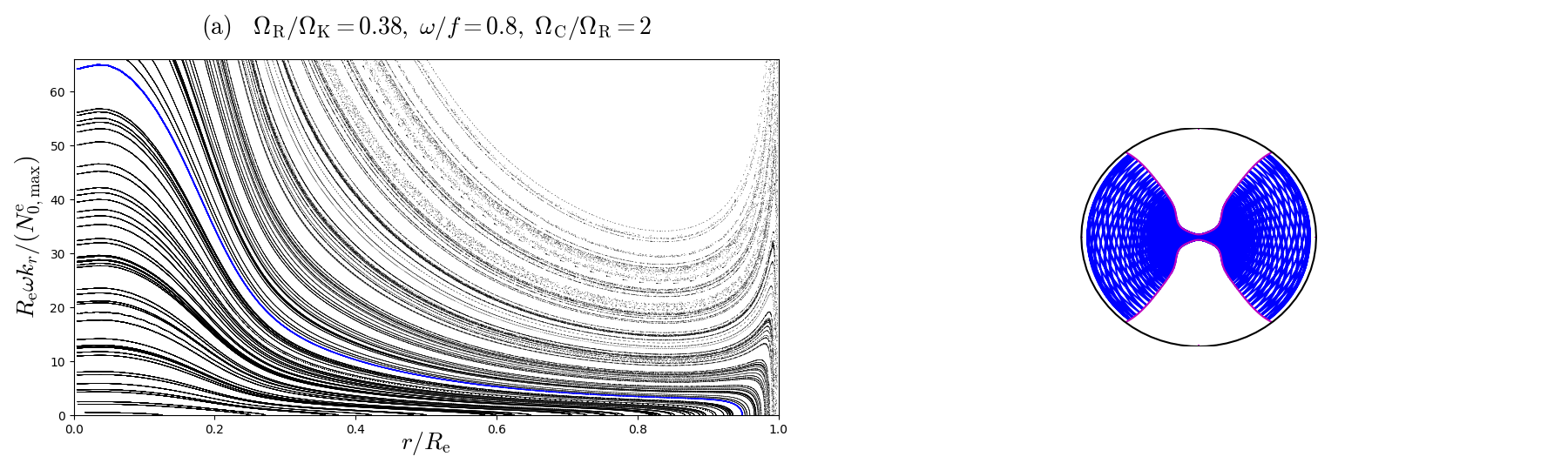}}}
    \resizebox{\hsize}{!}{\subfloat{\label{fig:2_trans}\includegraphics{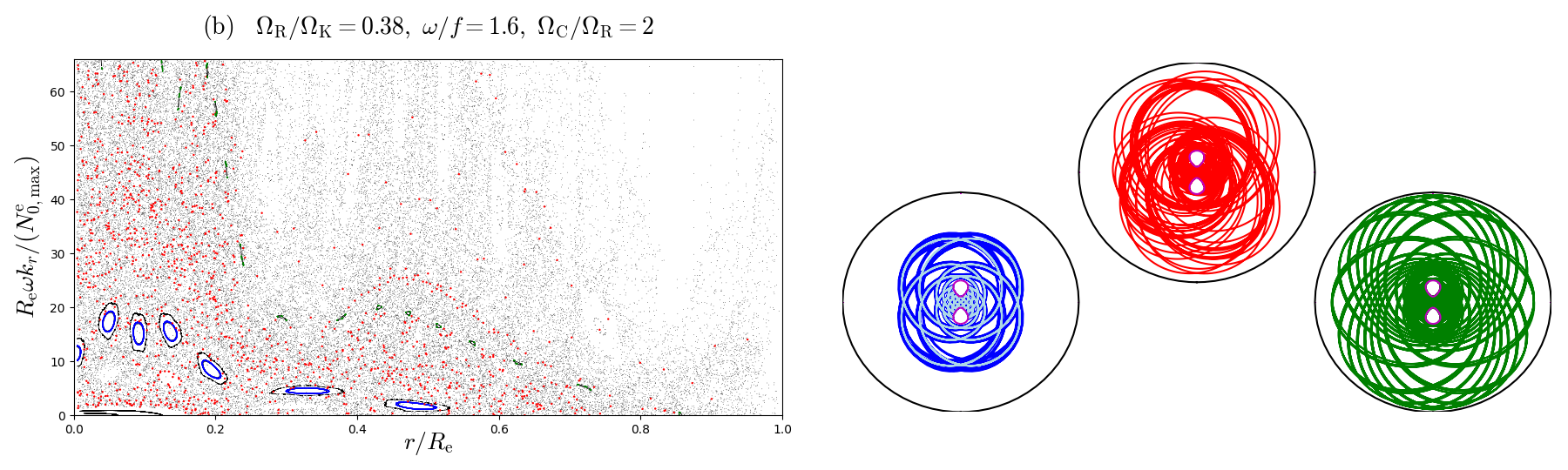}}}
    \resizebox{\hsize}{!}{\subfloat{\label{fig:2_sup}\includegraphics{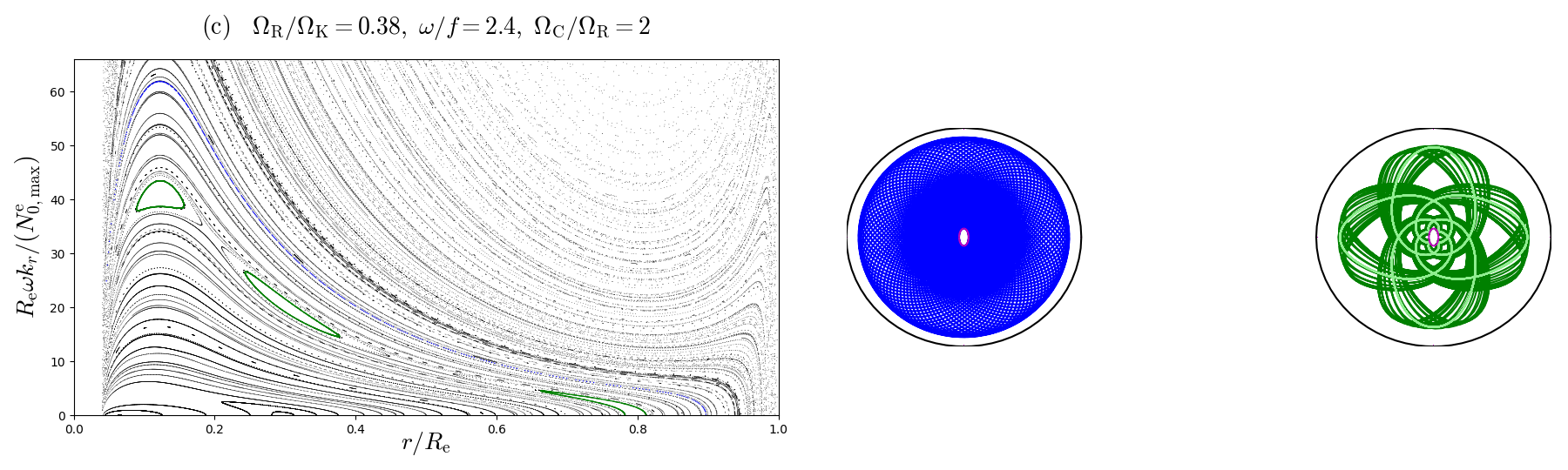}}}
    \caption{PSS (left) and examples of ray trajectories (right) for the three regimes identified in Fig.~\ref{fig:freq_oc2}: sub-, trans-, and super-inertial, respectively.
    Blue trajectories correspond to invariant tori (except for the second case, which corresponds to an island chain), green ones to island chains, and the red one to a chaotic zone.
    Lighter blue and green trajectories correspond to periodic orbits at the centre of the islands.
    Magenta lines are the limits of the domain of propagation.
    The imprints of the rays are shown on the PSS with colours corresponding to the rays.}
\end{figure*}

First, when $\omega<2\Omega_{\rm R}$, $\omega<2\Omega$ throughout the star, so according to Sect.~\ref{sec:domain}, rays avoid a region around the rotation axis and can propagate near the centre, as for sub-inertial rays in the uniformly rotating case.
We therefore call this the purely sub-inertial regime.
As illustrated in Fig.~\ref{fig:2_sub}, the structure of the phase space is dominated by invariant tori, as it is in the corresponding regime with uniform rotation.
Invariant tori are the only kind of structures present in integrable systems.

Second, when $\omega>2\Omega_{\rm C}$, $\omega>2\Omega$ everywhere in the star.
It follows that rays can propagate everywhere but near the centre, as in the uniformly rotating super-inertial case.
For that reason, we call it the purely super-inertial regime.
Figure~\ref{fig:2_sup} shows an example of PSS computed in this regime featuring the same kind of structure that was found in Paper~I: mainly invariant tori and island chains, such as those associated with rosette modes, which have been identified by \citet{Ballot12} and further described by \citet{TakataSaio, SaioTakata, Takata}.

\begin{figure}
    \resizebox{\hsize}{!}{\includegraphics{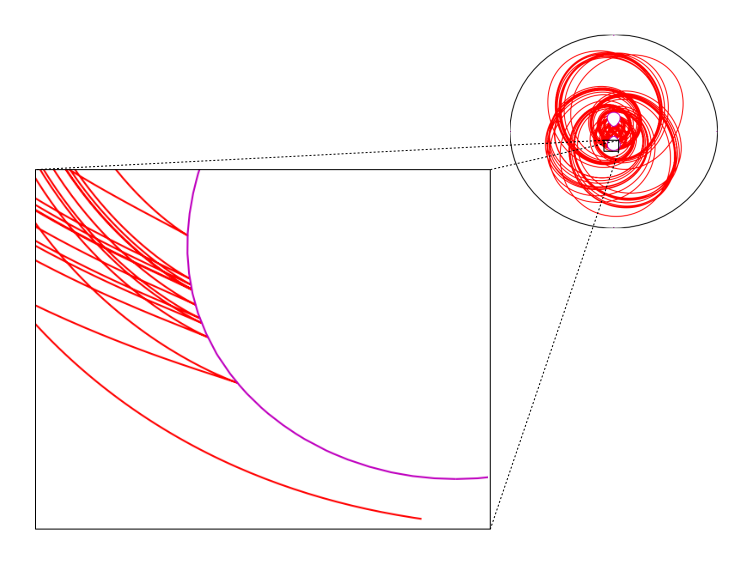}}
    \caption{Detail of a chaotic trans-inertial trajectory showing the transition between the sub-inertial and super-inertial behaviours.
    The ray bounces several times on the turning surface (sub-inertial, at the top) until it finally propagates past the rotation axis (super-inertial, at the bottom).}
    \label{fig:chaos}
\end{figure}

Third, when $2\Omega_{\rm R}<\omega<2\Omega_{\rm C}$, $\omega<2\Omega$ in most of the core, and $\omega>2\Omega$ in most of the radiative zone.
As a consequence, rays can propagate near the centre, and can propagate at any latitude in the radiative zone.
This new regime, that we call trans-inertial, is dominated by chaotic regions and a few island chains, as can be seen in Fig.~\ref{fig:2_trans}.
The presence of chaos may be explained by the fact that near the transition between the sub- and super-inertial regions, a small difference in position or momentum results in two completely different behaviours when approaching the rotation axis: either propagation if the ray is locally super-inertial, or reflection on the critical surface if the ray is locally sub-inertial.
This is illustrated in Fig.~\ref{fig:chaos}, which shows the detail of the red trajectory of Fig.~\ref{fig:2_trans}.

\begin{figure*}
    \resizebox{\hsize}{!}{\includegraphics{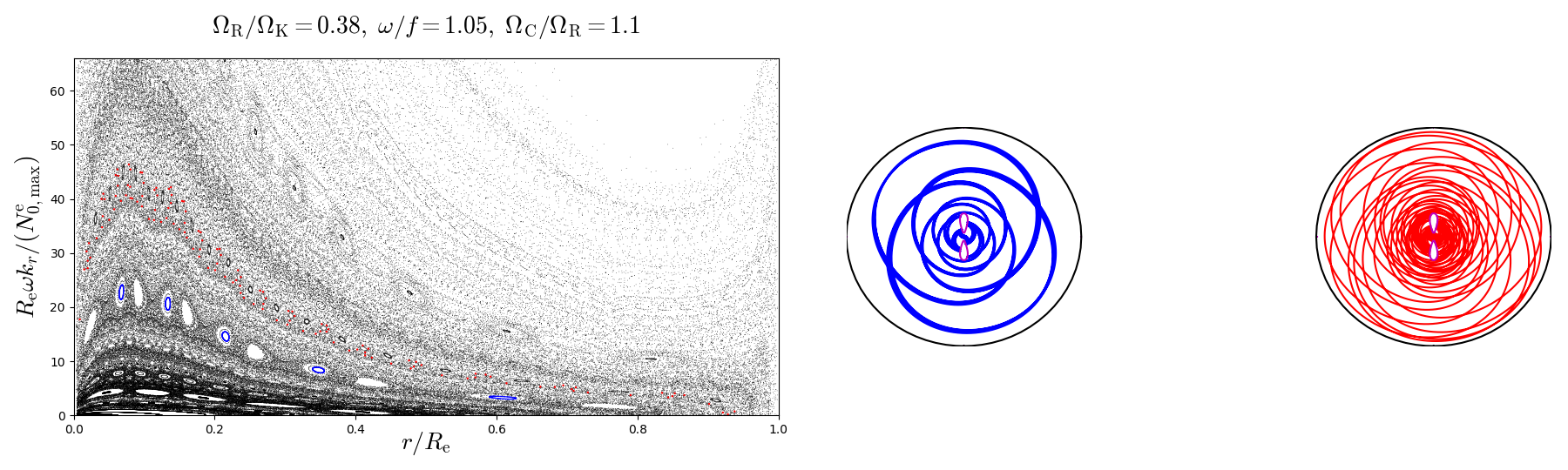}}
    \caption{PSS at $\Omega_{\rm C}/\Omega_{\rm R}=1.1$ and $\omega/(2\Omega_{\rm R})=1.05$ (left) and two trans-inertial rays: one belonging to an island chain (blue) and one belonging to a chaotic zone (red).}
    \label{fig:11_trans}
\end{figure*}

\begin{figure*}
    \resizebox{\hsize}{!}{\includegraphics{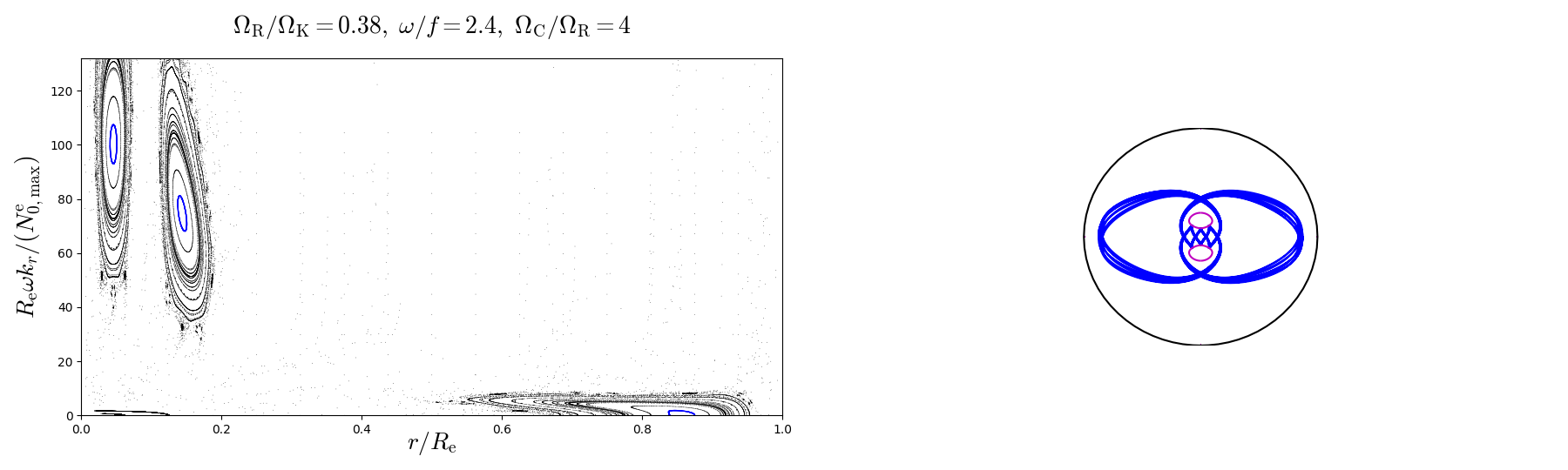}}
    \caption{PSS at $\Omega_{\rm C}/\Omega_{\rm R}=4$ and $\omega/(2\Omega_{\rm R})=2.4$ (left) and one trans-inertial ray belonging to an island chain (blue).}
    \label{fig:4_trans}
\end{figure*}

To determine if chaos persists in this regime even at low differential rotation, we computed a PSS with $\Omega_{\rm C}/\Omega_{\rm R}=1.1$ and $\omega/(2\Omega_{\rm R})=1.05$.
As can be seen on Fig.~\ref{fig:11_trans}, chaos is still dominant, with some island chains.
However, it seems to be the result of the juxtaposition of small chaotic zones, whereas we found a large chaotic zone when $\Omega_{\rm C}/\Omega_{\rm R}=2$.
We also computed a PSS with $\Omega_{\rm C}/\Omega_{\rm R}=4$ and $\omega/(2\Omega_{\rm R})=2.4$, which is shown in Fig.~\ref{fig:4_trans}.
One can see that the dynamics is dominated by chaos, except for a large island chain.

\subsubsection{Slow core}
\label{sec:cslow}

\begin{figure}
    \resizebox{\hsize}{!}{\includegraphics{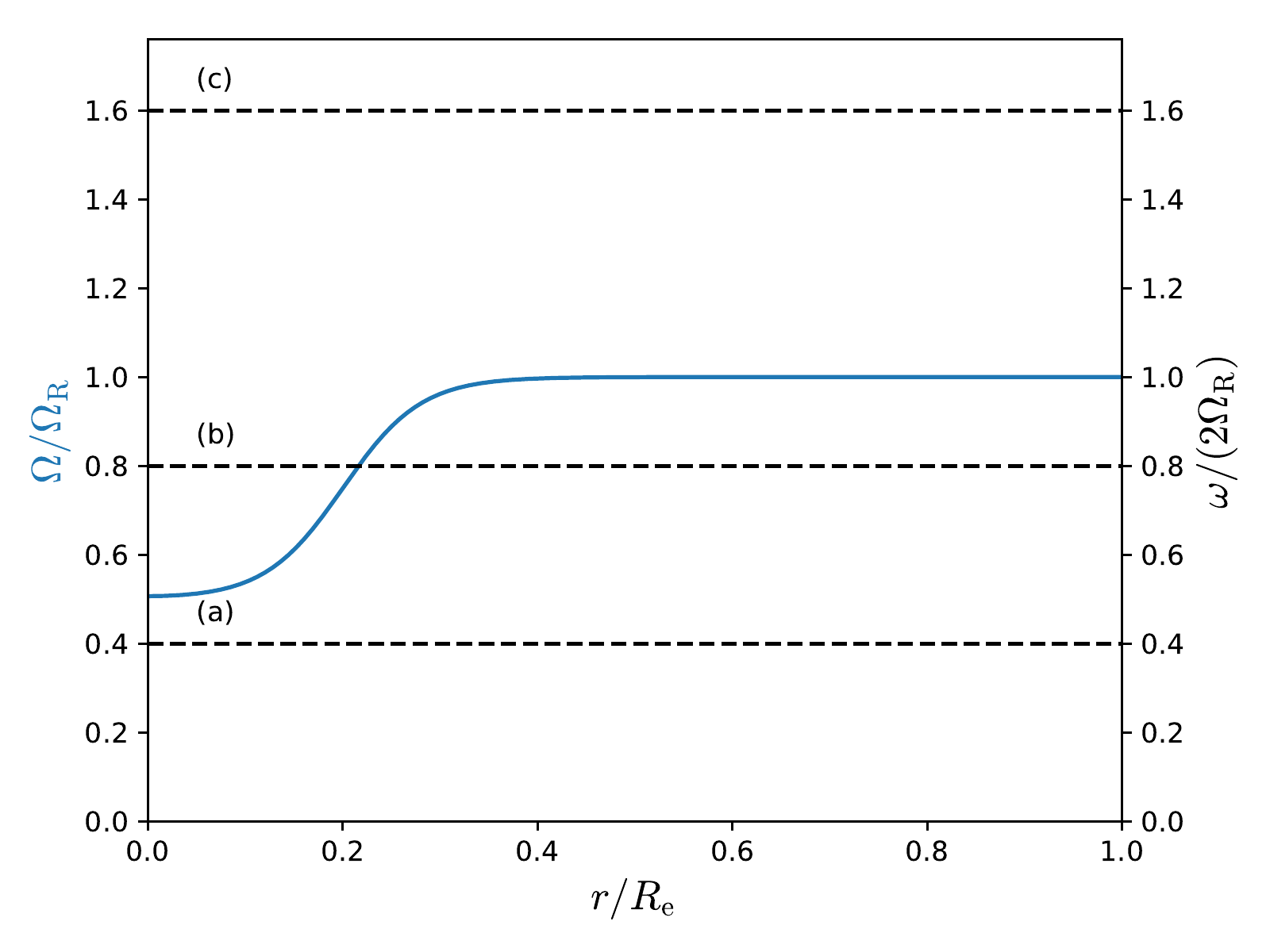}}
    \caption{Rotation profile with $\Omega_{\rm C}/\Omega_{\rm R}=0.5$, $r_{\rm C}=0.2$, and $\alpha_{\rm C}=25$ (blue solid line).
    The black dashed lines are the three frequencies for which we computed PSS: from bottom to top $\omega/(2\Omega_{\rm R})=0.4$, 0.8, and 1.6.
    They correspond to the sub-, trans- and super-inertial regimes, respectively.}
    \label{fig:freq_oc0.5}
\end{figure}

We now consider the case $\Omega_{\rm C}/\Omega_{\rm R}=0.5$.
Again, there are three different regimes, which are investigated through PSS computed at three frequencies shown in Fig.~\ref{fig:freq_oc0.5}.
\begin{figure*}
    \resizebox{\hsize}{!}{\subfloat{\label{fig:05_sub}\includegraphics{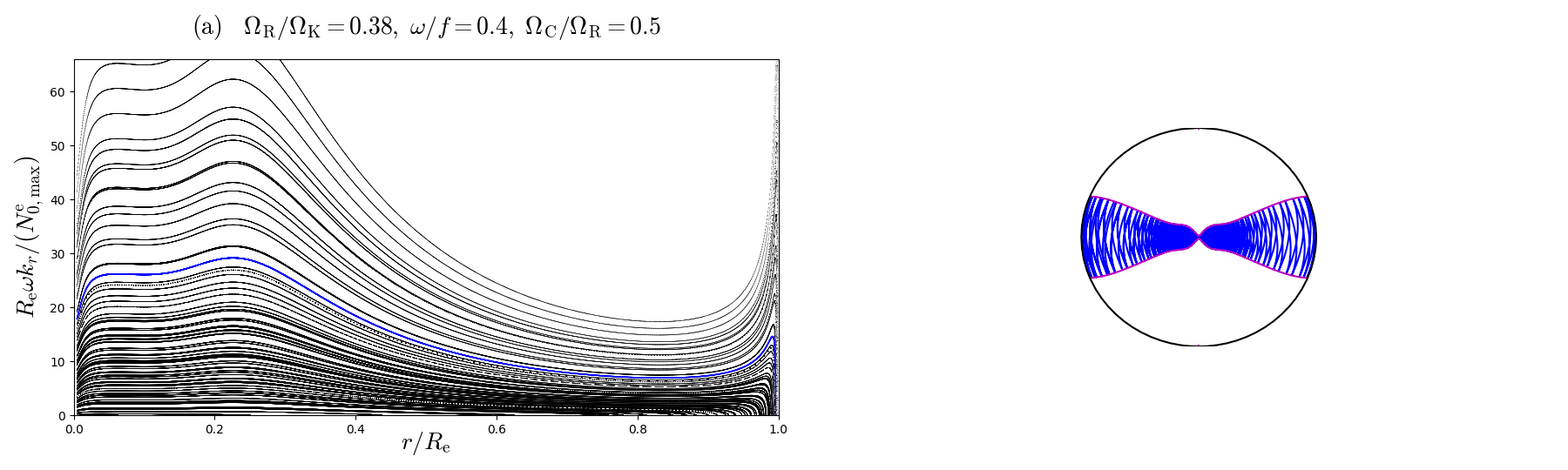}}}
    \resizebox{\hsize}{!}{\subfloat{\label{fig:05_trans}\includegraphics{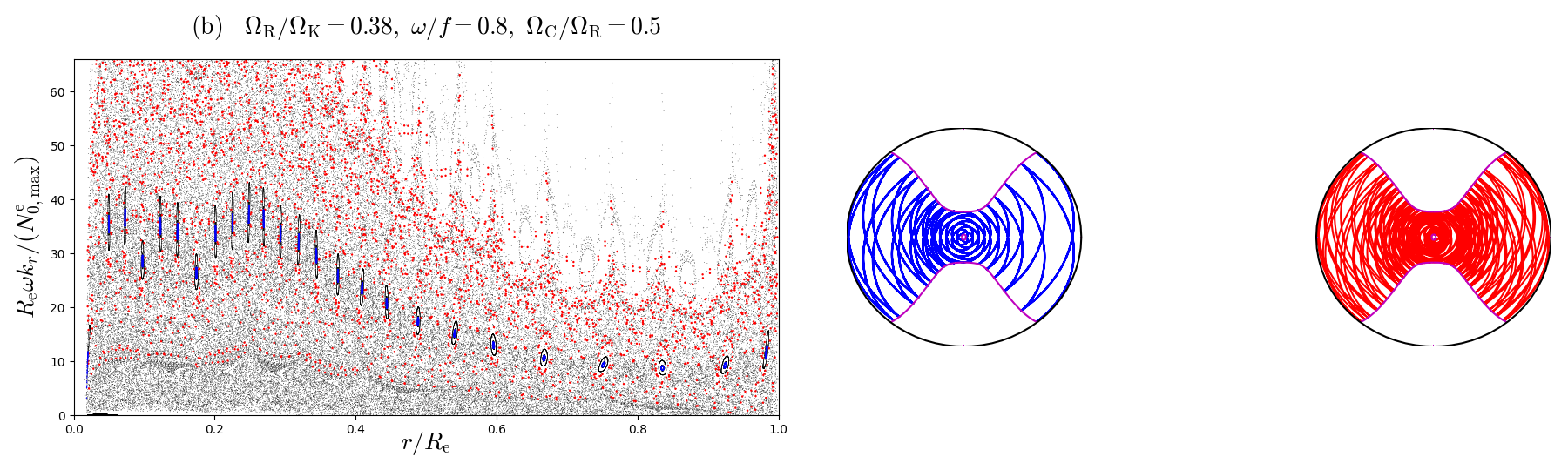}}}
    \resizebox{\hsize}{!}{\subfloat{\label{fig:05_sup}\includegraphics{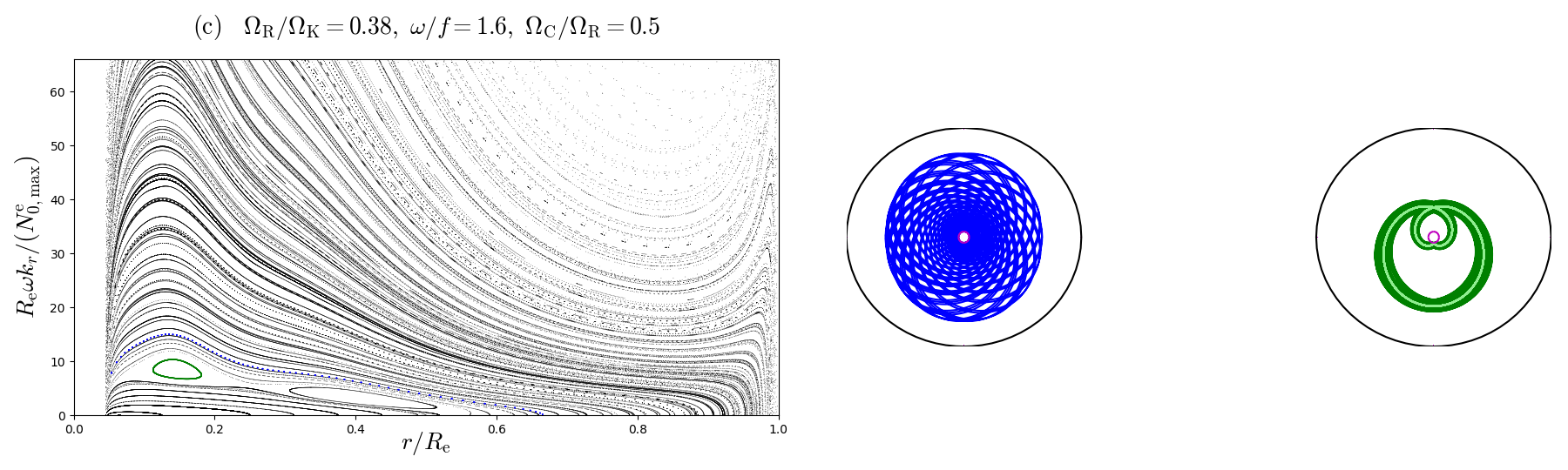}}}
    \caption{PSS (left) and examples of ray trajectories (right) for the three regimes shown in Fig.~\ref{fig:freq_oc0.5}: sub-, trans-, and super-inertial, respectively.
    Blue trajectories correspond to invariant tori (except for the second case, which corresponds to an island chain), green ones to island chains, and the red one to a chaotic zone.
    The lighter green trajectory corresponds to a periodic orbits at the centre of the island.}
\end{figure*}

In the sub-inertial regime, Fig.~\ref{fig:05_sub} shows that the dynamics is dominated by invariant tori, as in the case of a fast core.
This suggests that the near integrability at low frequencies that was observed with uniform rotation is still valid when considering a general radial differential rotation.
If this is indeed the case, it means that new seismic diagnoses similar to those obtained by \citet{PratMLBC} in the uniformly rotating case could be derived.

The super-inertial regime is also very similar to what we find for the case of the fast core, as can be seen in Fig.~\ref{fig:05_sup}.
In both sub- and super-inertial regimes, the main difference with the case of the fast core is the shape of the envelope.
As in the case of a fast core, the trans-inertial regime is largely dominated by chaos, with some stability islands.
However, the propagation domain, with avoided regions both at the centre (although it is rather small) and around the rotation axis, is different from the one obtained with a fast core.
This is illustrated in Fig.~\ref{fig:05_trans}.

\subsection{Latitudinal differential rotation in the envelope}
\label{sec:env}

In this section we investigate the effect of latitudinal differential rotation in the envelope on the structure of the phase space.
Such differential rotation takes place in convective envelopes of low-mass stars, as mentioned earlier, but also possibly in radiative envelopes of massive stars \citep[see e.g.][]{Rieutord13}.
Besides, at the interface between the envelope and the bulk of the radiative zone the latitudinal differential rotation of the envelope generates locally a potentially strong radial differential rotation, as it is in the solar tachocline.

\begin{figure}
    \resizebox{\hsize}{!}{\includegraphics{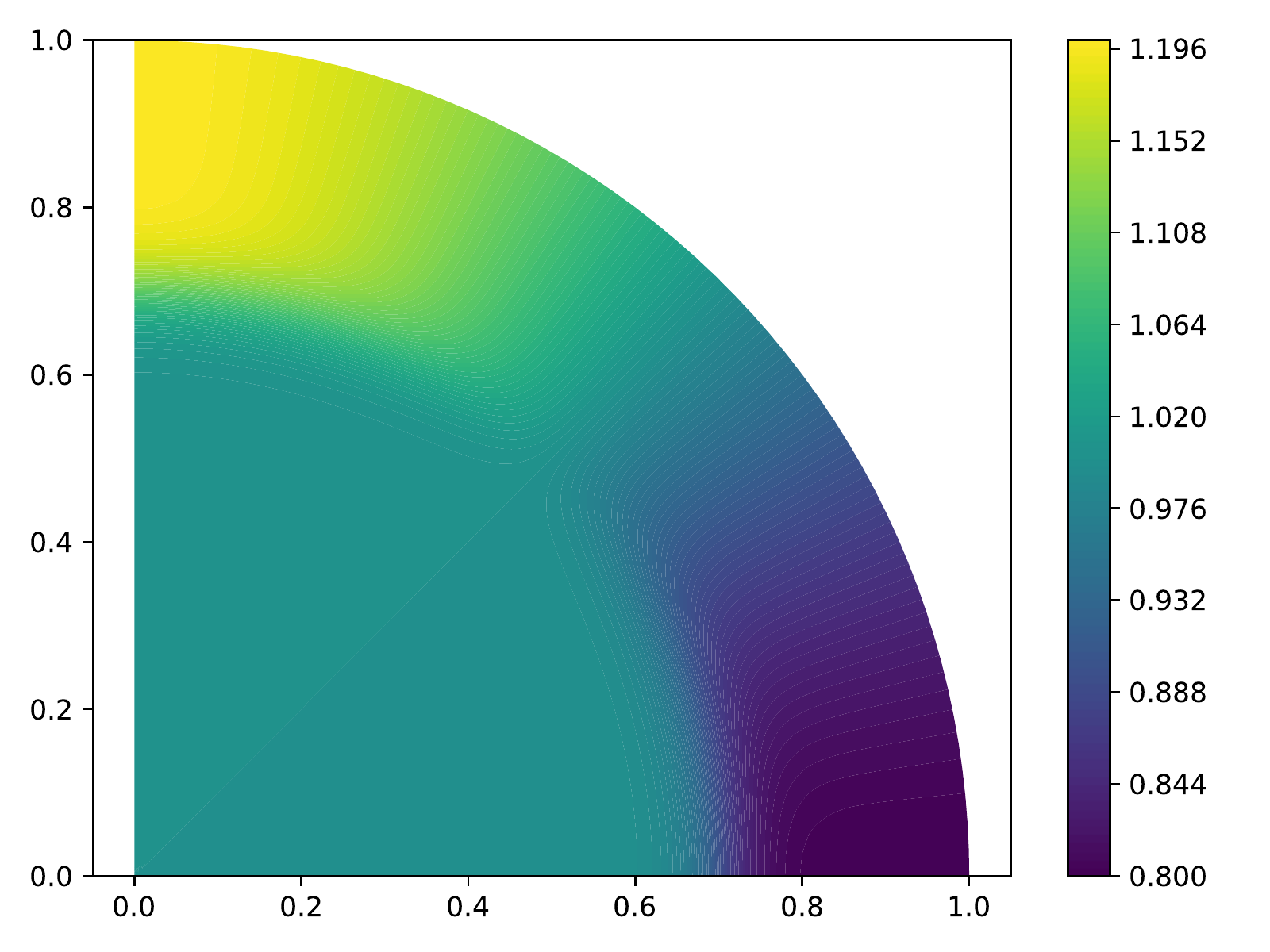}}
    \caption{Rotation profile with $\Omega_{\rm D}/\Omega_{\rm R}=0.2$, $\alpha_{\rm E}=40$, $r_{\rm E}=0.7$.}
    \label{fig:rot_lat}
\end{figure}

For simplicity, we assume a zero radial differential rotation near the core ($\Omega_{\rm C}=\Omega_{\rm R}$).
An example of rotation profile used in this section is given in Fig.~\ref{fig:rot_lat}.

\begin{figure}
    \resizebox{\hsize}{!}{\includegraphics{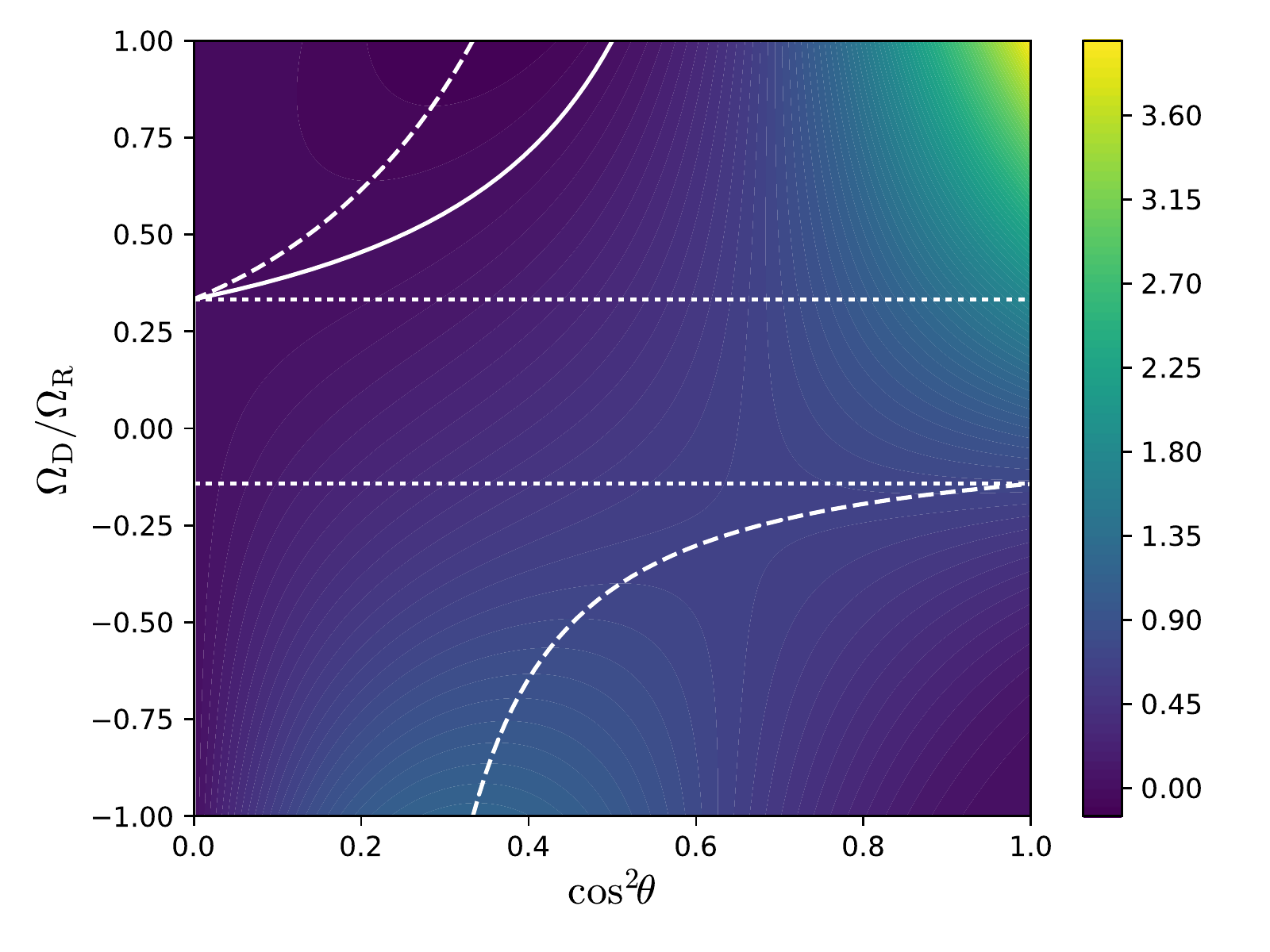}}
    \caption{Value of $\kappa^2=f\cos\theta(f\cos\theta+Q_\theta)$ as a function of $\eta=\Omega_{\rm D}/\Omega_{\rm R}$ and $\cos^2\theta$ in the envelope. The solid line corresponds to the contour $\kappa^2=0$. The dashed lines denote the position of the extremum of $\kappa^2$ as a funtion of $\theta$ at fixed $\eta$. The dotted lines delimits the region between $\eta=-1/7$ and $\eta=1/3$ where such an extremum does not exist.}
    \label{fig:crit_lat}
\end{figure}

As mentioned in Sect.~\ref{sec:domain}, the existence and the number of critical surfaces in latitude depends on the value of $f\cos\Theta(f\cos\Theta+Q_\perp)$ with respect to the angular frequency $\omega$.
To simplify the discussion, we neglect here the centrifugal deformation.
We thus consider the quantity $\kappa$ defined by $\kappa^2=f\cos\theta(f\cos\theta+Q_\theta)$, which is similar to a horizontal epicyclic frequency. 
Figure~\ref{fig:crit_lat} shows how $\kappa^2$ depends on the latitude and on the degree of differential rotation $\eta=\Omega_{\rm D}/\Omega_{\rm R}$.
First, $\kappa$ is always zero at the equator.

When $-1/7<\eta<1/3$, $\kappa^2$ is positive and monotonic,  $\kappa$ is real, and its maximum value $\kappa_{\rm max}$ is reached along the rotation axis.
When $\omega<\kappa_{\max}$, there is only one latitude at which $\omega=\kappa$, so there is only one critical latitude (sub-inertial regime).
The smaller the frequency, the closer the critical surface to the equatorial plane, as in the uniformly rotating case.
When $\omega>\kappa_{\rm max}$, there is no critical surface (super-inertial regime).

When $\eta>1/3$, $\kappa^2$ is negative near the equatorial plane and becomes positive and monotonic closer to the rotation axis, where it has a maximum value.
As in the previous case, we can define sub- and super-inertial regimes.
The main difference is that when the frequency tends to zero, the critical surface does not tend to the equatorial plane, but to another limit surface.

When $\eta<-1/7$, $\kappa^2$ is positive, but not monotonic.
$\kappa$ reaches a maximum value at a certain latitude, then decrease and reaches a positive value $\kappa_{\rm p}$ along the rotation axis.
When $\omega<\kappa_{\rm p}$, there is only one critical surface.
When $\kappa_{\rm p}<\omega<\kappa_{\rm max}$, there are two critical surfaces, and waves can propagate near the equatorial plane and the rotation axis.
We name this regime mid-inertial.
When $\omega>\kappa_{\rm max}$, there is no critical surface.

\begin{figure}
    \resizebox{\hsize}{!}{\includegraphics{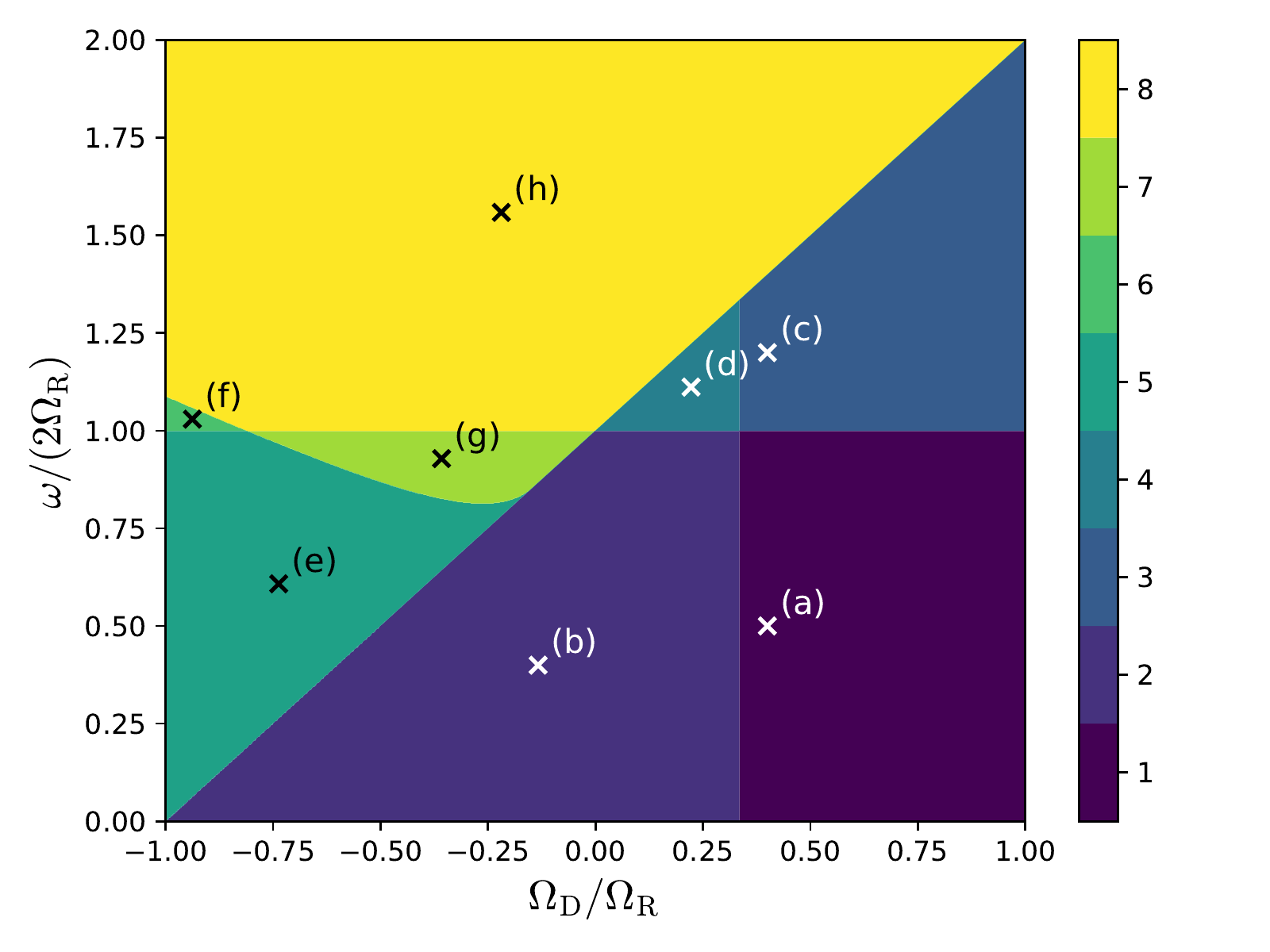}}
    \caption{Different regimes as a function of the frequency and the degree of differential rotation in the envelope.
    See main text for the description of these regimes.
    The markers correspond to the sets of parameters for which PSS have been computed.}
    \label{fig:regimes}
\end{figure}

\begin{figure*}
    \resizebox{\hsize}{!}{\subfloat{\label{fig:lata}\includegraphics{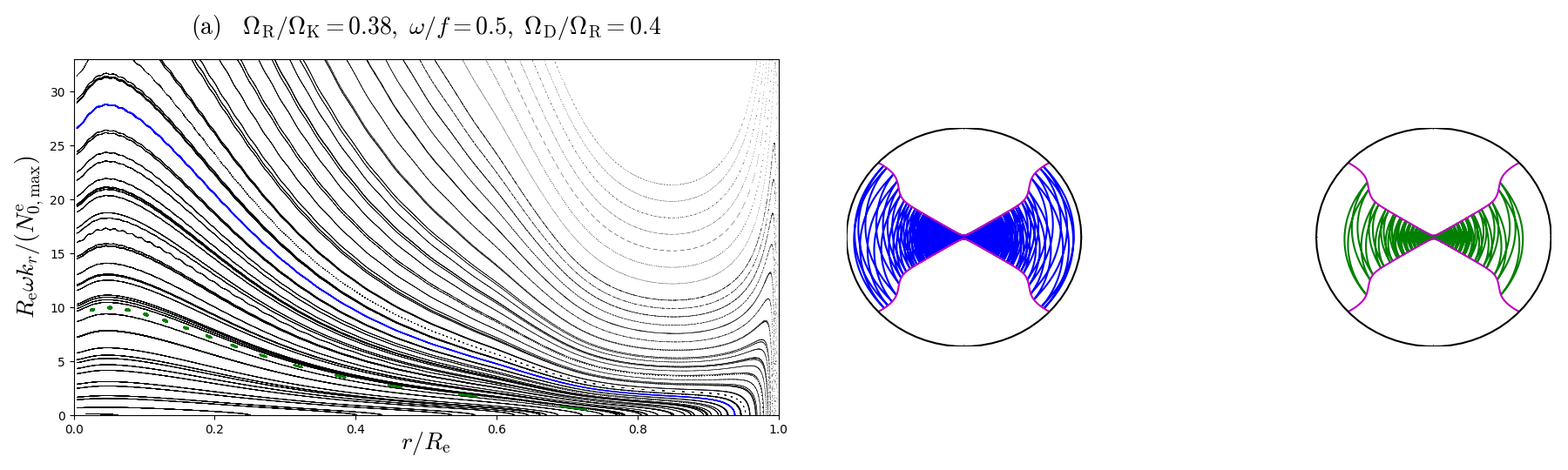}}}
    \resizebox{\hsize}{!}{\subfloat{\label{fig:latb}\includegraphics{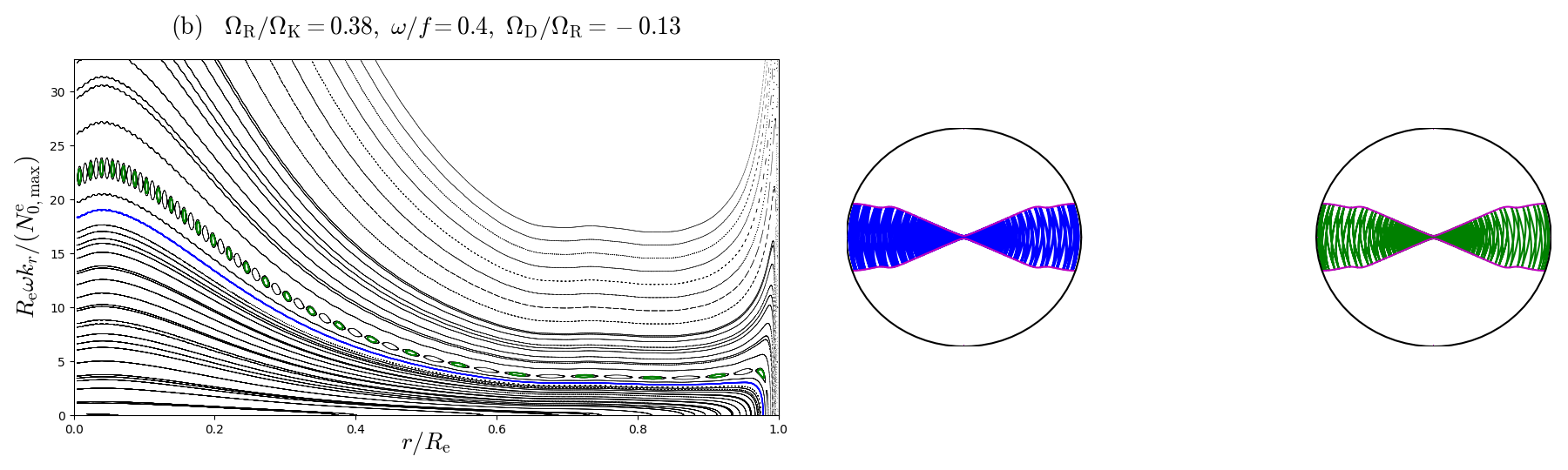}}}
    \resizebox{\hsize}{!}{\subfloat{\label{fig:latc}\includegraphics{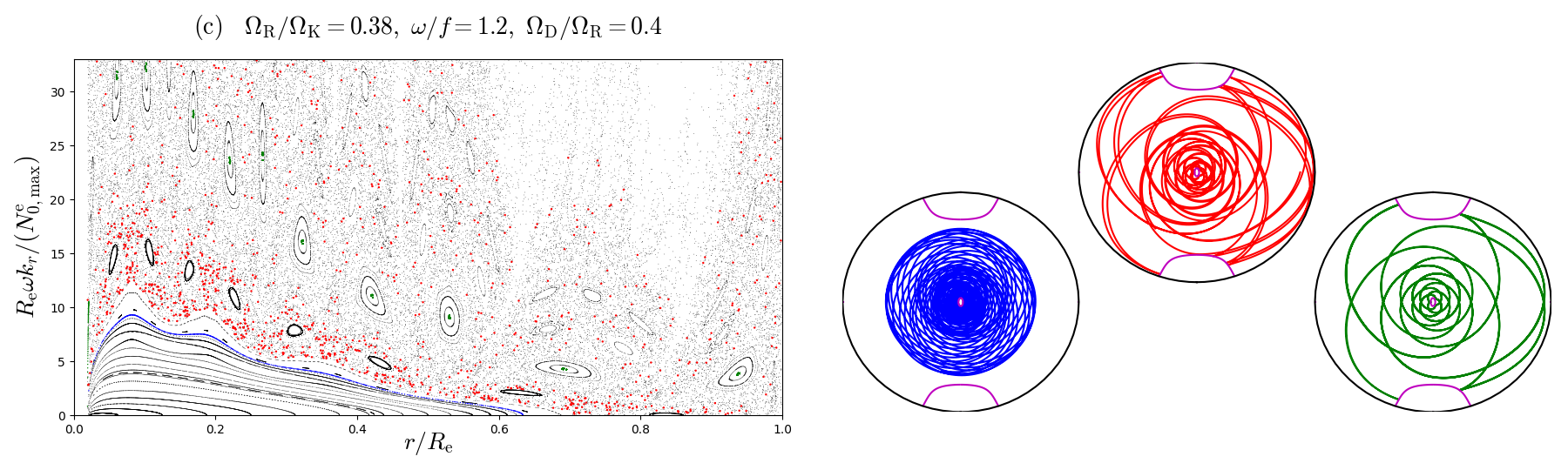}}}
    \resizebox{\hsize}{!}{\subfloat{\label{fig:latd}\includegraphics{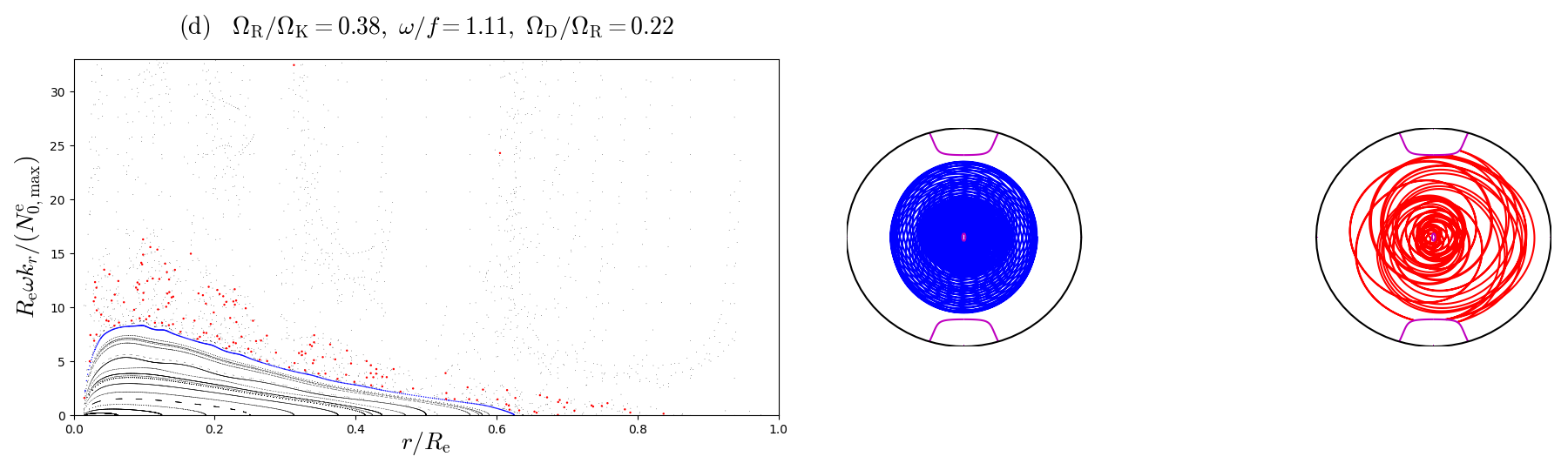}}}
    \caption{PSS (left) and examples of ray trajectories (right) for the four first regimes shown in Fig.~\ref{fig:regimes}.
Blue trajectories correspond to invariant tori, green ones to island chains, and red ones to chaotic zones.}
    \label{fig:4first}
\end{figure*}

In addition to the number of critical surfaces in the envelope, waves are either sub- or super-inertial in the inner region.
There are thus many different regimes, as illustrated in Fig.~\ref{fig:regimes}.
In the following, unless mentioned otherwise, we used $r_{\rm E}=0.7$ and $\alpha_{\rm E}=40$.
For Figs.~\ref{fig:lata} and~\ref{fig:latc}, which correspond to regimes with a strong anti-solar differential rotation, this would lead to the existence of regions that are unstable with respect to the Rayleigh-Taylor instability.
Since for real stars this instability would very rapidly change the background rotation towards a stable configuration, we chose to use $\alpha_{\rm E}=22$, which is a smoother transition, for the concerned calculations to avoid the existence of unstable regions.
However, the nature of the dynamics is not affected by this difference.
\addtocounter{figure}{-1}
\begin{figure*}
    \addtocounter{subfigure}{4}
    \resizebox{\hsize}{!}{\subfloat{\label{fig:late}\includegraphics{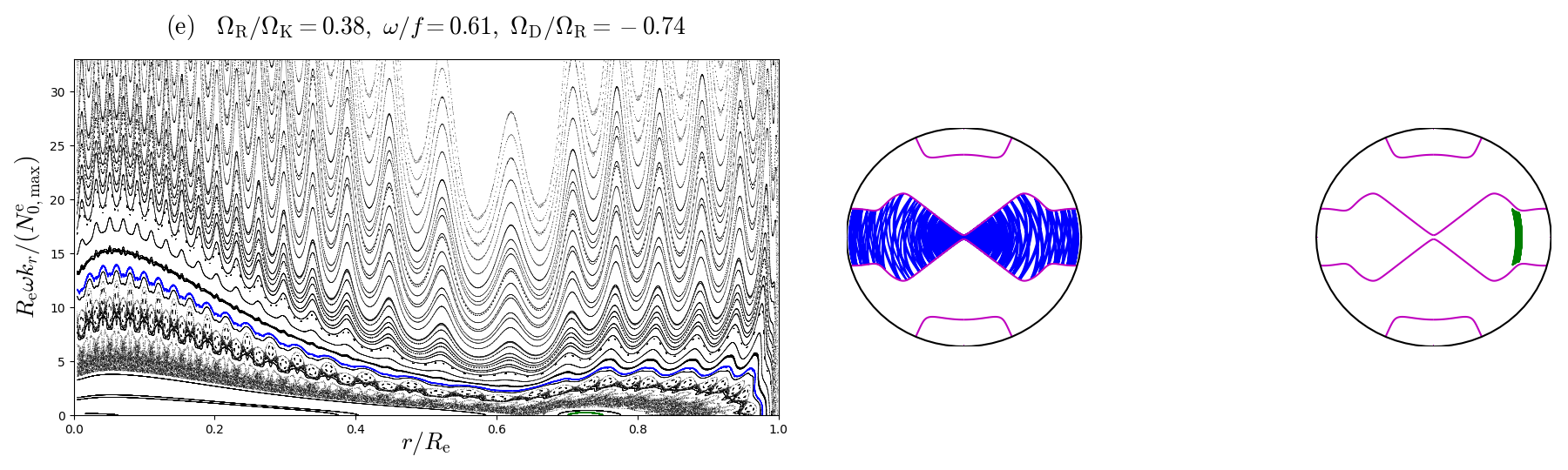}}}
    \resizebox{\hsize}{!}{\subfloat{\label{fig:latf}\includegraphics{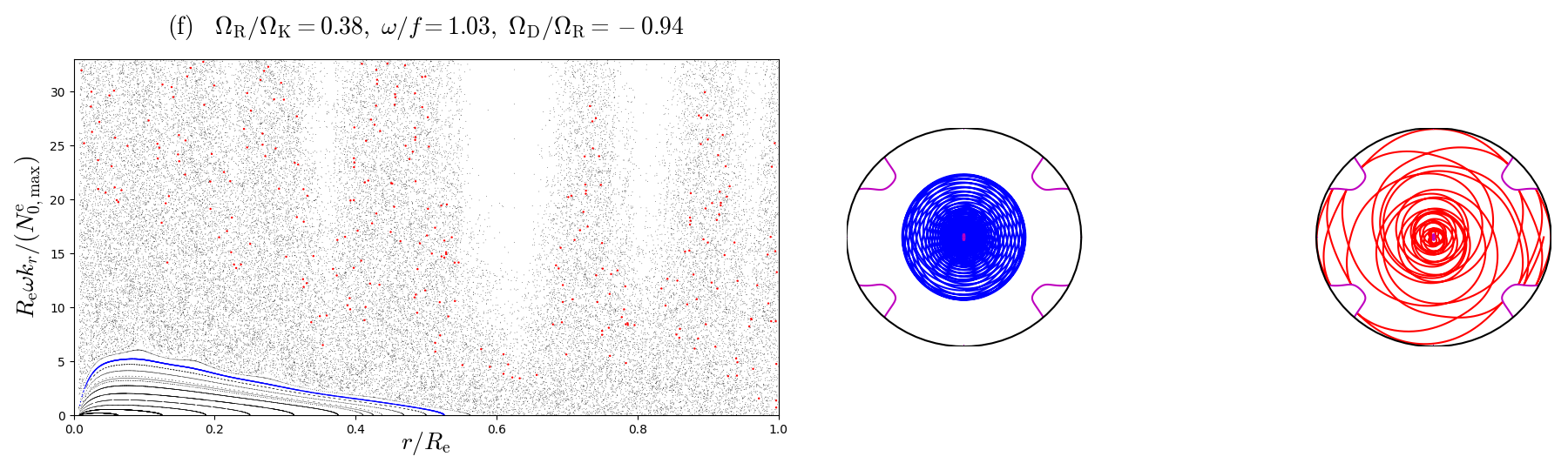}}}
    \resizebox{\hsize}{!}{\subfloat{\label{fig:latg}\includegraphics{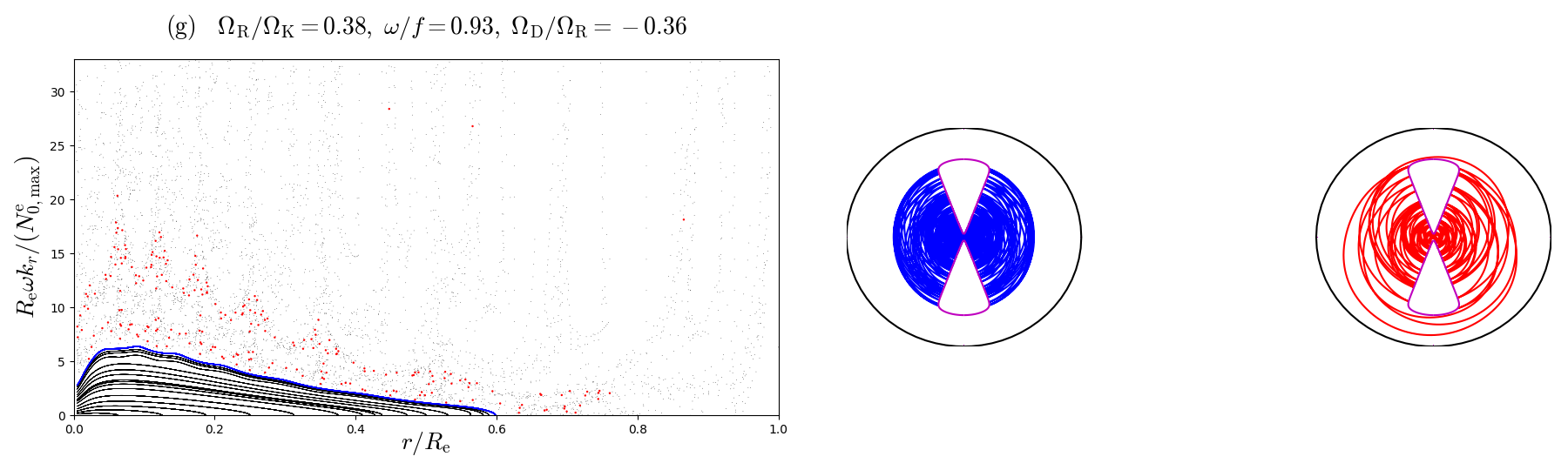}}}
    \resizebox{\hsize}{!}{\subfloat{\label{fig:lath}\includegraphics{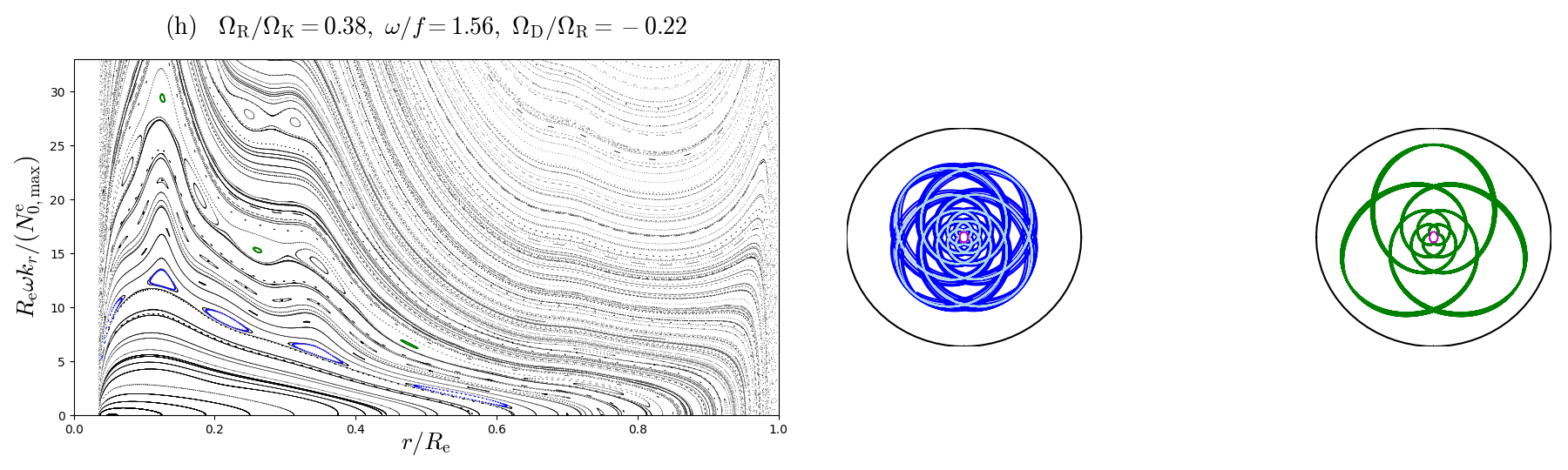}}}
    \caption{[continued] PSS (left) and examples of ray trajectories (right) for the four last regimes shown in Fig.~\ref{fig:regimes}.
Blue trajectories correspond to invariant tori (except for the last case, where it corresponds to an island chain, and where the light blue corresponds to the periodic orbit at the centre of the island), green ones to island chains, and red ones to chaotic zones.}
    \label{fig:4last}
\end{figure*}

Regimes 1 and 2 are purely sub-inertial.
The main difference between the two regimes is that at low frequency, waves are trapped very close to the equatorial plane in regime 2, and not in regime 1.
As shown in Figs.~\ref{fig:lata} and \ref{fig:latb}, these two regimes are largely dominated by invariant tori and tiny island chains, as expected for purely sub-inertial waves.

Regimes 3 and 4 are both sub-inertial in the envelope and super-inertial in the core, thus trans-inertial.
In Figs.~\ref{fig:latc} and \ref{fig:latd}, one can see that in contrast with regimes 1 and 2, there is not much difference in the domain of propagation, because the frequency is not small compared to the rotation frequency.
Both regimes are dominated by invariant tori when the maximum value of $k_r$ on the PSS for a given trajectory is low and by chaotic structures at higher values.
This can be explained by the fact that rays with low enough maximum values of $k_r$ do not propagate into the differentially rotating region, and thus behave as in the uniformly rotating case.
This was also the case in Sect.~\ref{sec:core}, but it was less visible due to the limited size of the core.
The main difference between regimes 3 and 4 is that several island chains of significant size are clearly visible inside the chaotic region in regime 3, but not in regime 4.

Regime 5 is sub-inertial in the core and mid-inertial in the envelope (two critical surfaces).
Because of the critical surface in the core, the two regions of the envelope where waves can propagate are not connected to each other.
As a consequence, the domain of propagation of waves that are trapped near the equatorial plane is similar to the one found in the purely sub-inertial regimes.
However, Fig.~\ref{fig:late} shows that in regime 5, invariant tori that dominate the phase space (as in the purely sub-inertial case) coexist with island chains and relatively small chaotic zones.
Moreover, the invariant tori seem to be less smooth than in the purely sub-inertial regime.
Although waves can also be trapped in the polar regions, such waves are not shown because the PSS based on the equatorial plane cannot capture their dynamics.

Regime 6 is super-inertial in the core and mid-inertial in the envelope.
In contrast to regime 5, here the two regions of the envelope where waves can propagate are connected to each other.
As illustrated in Fig.~\ref{fig:latf}, the structure of the phase space is very different depending on whether rays propagate in the envelope or not, as in regimes 3 and 4.

Regime 7 is sub-inertial in the core and super-inertial in the envelope.
Although the domain of propagation in this regime is very different for those in regimes 3, 4, and 6, the structure of the phase space is very similar, with different behaviours for rays that propagate into the envelope (and are chaotic) and for those that do not (and are nearly integrable).
This can be seen in Fig.~\ref{fig:latg}.

Regime 8 is purely super-inertial.
Figure~\ref{fig:lath} shows that in this regime, the phase space is dominated by invariant tori and island chains, similarly to the purely super-inertial regime of Sect.~\ref{sec:core}.

\section{Conclusions}
\label{sec:discussion}

In this paper we generalised the work done in Paper~I by introducing the effect of a general differential rotation on the ray dynamics for gravito-inertial waves.
In contrast to previous studies with differential rotation, we considered here the full Coriolis acceleration (i.e. without the traditional approximation) and the full centrifugal acceleration.
Focusing on axisymmetric waves as a first step, we wrote the equations governing the rays dynamics and implemented them in a ray-tracing code.
We then numerically investigated the domain of propagation of rays and the nature of their dynamics in various regimes of differential rotation.

We find that differential rotation can generate a large variety of domains of propagation.
However, one can distinguish between three main regimes for the nature of the ray dynamics.
At low frequency, we observe a regime similar to the sub-inertial regime in the case of uniform rotation, where waves are trapped near the equatorial plane.
The dynamics is dominated by invariant tori, even though these structures are deformed with respect to the uniformly rotating case.
At high frequency, rays behave as in the super-inertial regime of the uniformly rotating case, with mostly invariant tori and island chains, and sometimes chaotic zones in a narrow range of frequency (in the vicinity of the inner maximum of the Brunt-V\"ais\"al\"a frequency, see Paper~I for more details).
Between these two regimes, we find a new regime, called trans-inertial.
This regime is characterised by a dynamics dominated by chaotic zones (sometimes with some island chains) for rays that propagate into differentially rotating regions and by invariant tori for rays that stay in regions with negligible differential rotation.

The properties of the modes, which result from the superposition of positively interfering rays, can be inferred from semi-classical quantisation methods and concepts mostly developed in the domain of quantum physics (see for example Sect.~5 of Paper I).
Accordingly, the spectrum of axisymmetric gravito-inertial modes in fast rotators should have the following properties:
(i) at low frequency, mostly regular modes;
(ii) at high frequency, mostly regular modes and island modes, and a few chaotic modes;
and (iii) in between, mostly chaotic modes, in a range that depends on the amount of differential rotation.
The complete picture should nevertheless be more complicated, since non-axisymmetric modes may behave differently.

The large variety of domains of propagation generated by differential rotation means that gravito-inertial waves can probe many different resonant cavities in differentially rotating stars.
The properties of these cavities have a key influence on the visibility of modes and on the transport of angular momentum by waves, which need to extract angular momentum in excitation regions and deposit it in damping regions \citep[see e.g.][]{Mathis09}.

As already mentioned, the purely super-inertial regime is dominated by invariant tori.
This suggests the existence of a nearby integrable system that could provide us with seismic diagnoses for low-frequency gravito-inertial modes, as done in the case of uniform rotation in \citet{PratMLBC}.

In the present work we did not use self-consistent rotating stellar models, as we neglected the effect of differential rotation on stellar structure.
Determining to what extent our results still apply for more realistic background models will require studying the ray dynamics in baroclinic and centrifugally deformed stellar models such as ESTER models \citep{Rieutord16}.
Another possible extension of the present work would be to consider the effect of differential rotation on acoustic rays.
As the Coriolis force has a negligible effect on high-frequency acoustic waves, we expect the main effect to come from the way differential rotation modifies the stellar background model.
Finally, our predictions on the properties of gravito-inertial modes based on ray dynamics should also be compared with numerically computed modes, using bi-dimensional codes such as TOP \citep{Reese06} or ACOR \citep{Ouazzani12}, knowing that for acoustic modes \citep{LG09, Pasek11, Pasek12} and some gravito-inertial modes \citep{Ballot12}, such comparisons have so far been successful.

The ray dynamics can also be used to interpret the huge amount of data produced by global 3D time-dependent simulations of the excitation, propagation, and dissipation of non-linear waves \citep{Alvan14, Alvan15}.

In the near future, we plan to generalise the present work to non-axisymmetric waves.
This will require a significant amount of work, since the eikonal equation \eqref{eq:eik_full} we derived for non-axisymmetric waves is more complex than the one for axisymmetric waves, and the Doppler-shifting of waves needs to be taken into account.
Since the transport of angular momentum by waves comes from the difference in the excitation and the damping between prograde and retrograde waves, it is crucial to understand the physics of non-axisymmetric waves.
Eventually, it will also be necessary to include the effect of dissipative processes in the present formalism to describe the transport generated by the waves.

Another physical ingredient that has been neglected in this study is the magnetic field.
In addition to modifying the background structure of the stars, it also affects the propagation of gravito-inertial waves, which become magneto-gravito-inertial waves \citep{MathisdeBrye11}, and the transport of angular momentum due to these waves \citep{MathisdeBrye12}.

\begin{acknowledgements}
    V.P., S.M., and K.A. acknowledge support from the European Research Council through ERC grant SPIRE 647383.
    V.P., F.L., and J.B. acknowledge the International Space Science Institute (ISSI) for supporting the SoFAR international team\footnote{\url{http://www.issi.unibe.ch/teams/sofar/}}.
    The authors acknowledge funding by SpaceInn, PNPS (CNRS/INSU), and CNES CoRoT/\textit{Kepler} and PLATO grants at DAp and IRAP.
    The authors also thank the referee for giving useful comments that have allowed us to improve the article.
\end{acknowledgements}

\bibliographystyle{aa}
\bibliography{refs}

\appendix
\onecolumn

\section{Derivation of the wave equation}
\label{sec:deriv}

In this appendix, we want to derive a wave equation for one variable only from the perturbation equations~\eqref{eq:cont}-\eqref{eq:iso}.

First, we assume time-harmonicity, and look for solutions of the form $A(\vec x) = \Re[\hat A(r,\theta) e^{i(m\varphi-\omega t)}]$.
Density fluctuations can be expressed in terms of pressure and velocity fluctuations using Eq.~\eqref{eq:iso}:
\begin{equation}
    \label{eq:rhop_raw}
    \hat\rho = \frac{\hat P}{{c_{\rm s}}^2} - i\frac{\vec{\hat u}}{\hat\omega}\cdot\left(\frac{\grad\rho_0}{\rho_0}-\frac{\grad P_0}{{c_{\rm s}}^2\rho_0}\right),
\end{equation}
where $\vec{\hat u}= \rho_0\vec{\hat v}$ and $\hat\omega=\omega-m\Omega$.
It is possible to link the right-hand side of the dot product to the Brunt-Väisälä frequency defined by
\begin{equation}
    \label{eq:bv}
    {N_0}^2 = \vec g_0\cdot\left(\frac{\grad\rho_0}{\rho_0}-\frac{\grad P_0}{\Gamma_1 P_0}\right),
\end{equation}
where $\Gamma_1$ is the first adiabatic exponent of the fluid.
To do so, we use the relation
\begin{equation}
    \label{eq:baroc}
    \frac{\grad\rho_0}{\rho_0} = \left(\frac{\grad\rho_0}{\rho_0}\cdot\vec g_0\right)\frac{\vec g_0}{{g_0}^2} + \frac{(\grad\wedge\vec g_0)\wedge\vec g_0}{{g_0}^2},
\end{equation}
which derives from Eq.~\eqref{eq:geff}.
Using the definition of the effective gravity given in Eq.~\eqref{eq:defg}, Eq.~\eqref{eq:rhop_raw} becomes
\begin{equation}
    \hat\rho = \frac{\hat P}{{c_{\rm s}}^2} + i\frac{\vec{\hat u}}{\hat\omega g_0}\cdot({N_0}^2\vec e_\parallel + fQ_z\vec e_\perp),
\end{equation}
where $\vec e_\parallel = -\vec g_0/g_0$ and $\vec e_\perp$ is the unit vector orthogonal to $\vec e_\parallel$ such that $(\vec e_\parallel, \vec e_\perp, \vec e_\varphi)$ is a direct orthonormal basis (see Fig.~\ref{fig:coords}). 

The previous equation can now be used to eliminate $\rho$ in Eqs~\eqref{eq:cont} and \eqref{eq:mom_pert}:
\begin{align}
    -i\hat\omega\frac{\hat P}{{c_{\rm s}}^2}+\frac{\vec{\hat u}}{g_0}\cdot({N_0}^2\vec e_\parallel+fQ_z\vec e_\perp) + \grad\cdot\vec{\hat u}+\frac{imu_\varphi}{r\sin\theta}   &= 0,   \label{eq:cont_mat}\\
    \vec{\mathcal{M}}\cdot\vec{\hat u}                                                                                                                                          &= -\grad \hat P  -\frac{im\hat P\vec e_\varphi}{r\sin\theta} + \frac{\hat P\vec g_0}{{c_{\rm s}}^2},    \label{eq:mom_mat}
\end{align}
where the tensor $\vec{\mathcal{M}}$ is described in the basis $(\vec e_\parallel, \vec e_\perp, \vec e_\varphi)$ by the matrix
\begin{equation}
    \begin{bmatrix}
        -\dfrac{i}{\hat\omega}(\hat\omega^2-{N_0}^2) &   \dfrac{i}{\hat\omega}fQ_z    &   -f\sin\Theta    \\
        0                                           &   -i\hat\omega                &   -f\cos\Theta    \\
        f\sin\Theta+Q_\parallel                     &   f\cos\Theta+Q_\perp         &   -i\hat\omega
    \end{bmatrix},
\end{equation}
and $\Theta$ is the angle between $\vec e_z$ and $\vec e_\parallel$, which is equal to the co-latitude $\theta$ when the centrifugal deformation is neglected.
When $\vec{\mathcal{M}}$ is invertible, Eq.~\eqref{eq:mom_mat} can be used to express $\vec{\hat u}$ as a function of $\hat P$ and its gradient:
\begin{equation}
    \label{eq:up}
    \vec{\hat u}=\vec{\mathcal{M}}^{-1}\cdot\left( -\grad \hat P -\frac{im\hat P\vec e_\varphi}{r\sin\theta} + \frac{\hat P\vec g_0}{{c_{\rm s}}^2}\right),
\end{equation}
where
\begin{equation}
    \vec{\mathcal{M}}^{-1} = -\frac{1}{\Gamma}
    \begin{bmatrix}
        i\hat\omega[\hat\omega^2-f\cos\Theta(f\cos\Theta+Q_\perp)]  &   i\hat\omega f\cos\Theta(f\sin\Theta+Q_\parallel)                            &   -f(\hat\omega^2\sin\Theta+fQ_z\cos\Theta)  \\
        i\hat\omega f\cos\Theta(f\sin\Theta+Q_\parallel)            &   i\hat\omega[\hat\omega^2-{N_0}^2-f\sin\Theta(f\sin\Theta+Q_\parallel)]      &   -f\cos\Theta(\hat\omega^2-{N_0}^2)  \\
        \hat\omega^2(f\sin\Theta+Q_\parallel)                       &   (\hat\omega^2-{N_0}^2)(f\cos\Theta+Q_\perp)+fQ_z(f\sin\Theta+Q_\parallel)   &   i\hat\omega(\hat\omega^2-{N_0}^2)
    \end{bmatrix}
\end{equation}
and $\Gamma$ is defined in Eq.~\eqref{eq:gamma}.
Finally, the combination of Eqs.~\eqref{eq:cont_mat} and \eqref{eq:up} leads to a single equation for $\hat P$ only:
\begin{equation}
    -i\hat\omega\frac{\hat P}{{c_{\rm s}}^2}+\left(\frac{{N_0}^2\vec e_\parallel+fQ_z\vec e_\perp}{g_0}+\frac{im\vec e_\varphi}{r\sin\theta}\right)\cdot\left[\vec{\mathcal{M}}^{-1}\cdot\left( -\grad \hat P -\frac{im\hat P\vec e_\varphi}{r\sin\theta} + \frac{\hat P\vec g_0}{{c_{\rm s}}^2}\right)\right] + \grad\cdot\left[\vec{\mathcal{M}}^{-1}\cdot\left( -\grad \hat P -\frac{im\hat P\vec e_\varphi}{r\sin\theta} + \frac{\hat P\vec g_0}{{c_{\rm s}}^2}\right)\right]   = 0.
\end{equation}
Equivalently, using the tensor identity $\grad\cdot(\vec{\mathcal A}\cdot \vec b) = (\grad\cdot\vec{\mathcal A})\cdot\vec b + \vec{\mathcal{A}}^T:\grad\vec b$, where $\vec{\mathcal A}$ is a tensor, yields the Poincar\'e equation
\begin{equation}
    \label{eq:wave}
    \vec{\mathcal{A}}:\grad\grad\hat P + \vec B\cdot\grad\hat P + C\hat P = 0,
\end{equation}
where
\begin{align}
    \vec{\mathcal{A}}   &=  -(\vec{\mathcal{M}}^{-1})^T,    \\
    \vec B  &=  -\left(\frac{{N_0}^2\vec e_\parallel+fQ_z\vec e_\perp}{g_0}+\frac{im\vec e_\varphi}{r\sin\theta}\right)\cdot\vec{\mathcal{M}}^{-1} - \grad\cdot(\vec{\mathcal{M}}^{-1}) + \left(\frac{\vec g_0}{{c_{\rm s}}^2}-\frac{im\vec e_\varphi}{r\sin\theta}\right)\cdot(\vec{\mathcal{M}}^{-1})^T,   \label{eq:b}   \\
    C       &=  -\frac{i\hat\omega}{{c_{\rm s}}^2} + \left(\frac{{N_0}^2\vec e_\parallel+fQ_z\vec e_\perp}{g_0}+\frac{im\vec e_\varphi}{r\sin\theta}\right)\cdot\vec{\mathcal{M}}^{-1}\cdot\left(\frac{\vec g_0}{{c_{\rm s}}^2}-\frac{im\vec e_\varphi}{r\sin\theta}\right) + \grad\cdot\left[\vec{\mathcal{M}}^{-1}\cdot\left(\frac{\vec g_0}{{c_{\rm s}}^2}-\frac{im\vec e_\varphi}{r\sin\theta}\right)\right]. \label{eq:c}
\end{align}

\section{Determination of the surface term}
\label{sec:surface}

Close to the surface, $c_{\rm s}$ becomes very small.
Let us consider here the case where $N_0$ remains finite near the surface.
The dominant term of $\vec B$ is
\begin{equation}
    \vec B_0 = \frac{\vec g_0}{{c_{\rm s}}^2}\cdot(\vec{\mathcal{M}}^{-1})^T,
\end{equation}
and the one of $C$ is
\begin{equation}
    C_0 = \vec g_0\cdot(\vec{\mathcal{M}}^{-1})^T\cdot\grad\left(\frac{1}{{c_{\rm s}}^2}\right).
\end{equation}
Using the relation ${c_{\rm s}}^2 = \Gamma_1P_0/\rho_0$, along with Eqs.~\eqref{eq:bv} and~\eqref{eq:baroc}, one can write
\begin{equation}
    \vec\grad({c_{\rm s}}^2) = \beta\vec g_0 + {c_{\rm s}}^2\left[\frac{\vec\nabla\Gamma_1}{\Gamma_1} - \frac{(\vec\nabla\wedge\vec g_0)\wedge\vec g_0}{{g_0}^2}\right],
\end{equation}
where $\beta=\Gamma_1-1-\alpha$ and $\alpha={c_{\rm s}}^2{N_0}^2/{g_0}^2$, which tends to zero at the surface for convective envelopes and to a finite value for polytropic radiative envelopes.
Assuming that the term in brackets in the equation above remains finite and that $c_{\rm s}$ vanishes at the surface implies that near the surface $\grad({c_{\rm s}}^2) \simeq \beta\vec g_0$.
Thus,
\begin{equation}
    C_0 \simeq -\beta\frac{\vec g_0}{{c_{\rm s}}^4}\cdot(\vec{\mathcal{M}}^{-1})^T\cdot\vec g_0.
\end{equation}

Now, the wave equation~\eqref{eq:wave} is put in the so-called normal form to get rid of first order terms.
This is done by writing $\hat P = a\psi$ with a well-chosen function $a$.
If one retains only dominant terms of $\vec B$ and $C$, and using the fact that the double gradient is symmetric, Eq.~\eqref{eq:wave} becomes
\begin{equation}
    \label{eq:apsi}
    -a\vec{\mathcal{N}}:\grad\grad\psi + (-2\grad a\cdot\vec{\mathcal{N}}+a\vec B_0)\cdot\grad\psi + (-\vec{\mathcal{N}}:\grad\grad a+\vec B_0\cdot\grad a+C_0a)\psi = 0,
\end{equation}
where
\begin{equation}
    \vec{\mathcal{N}} = -\frac{1}{\Gamma}
    \begin{bmatrix}
        i\hat\omega[\hat\omega^2-f\cos\Theta(f\cos\Theta+Q_\perp)]  &   i\hat\omega f\cos\Theta(f\sin\Theta+Q_\parallel)                            &   \frac{1}{2}(\hat\omega^2Q_\parallel-f^2Q_z\cos\Theta)   \\
        i\hat\omega f\cos\Theta(f\sin\Theta+Q_\parallel)            &   i\hat\omega[\hat\omega^2-{N_0}^2-f\sin\Theta(f\sin\Theta+Q_\parallel)]      &   \frac{1}{2}[(\hat\omega^2-{N_0}^2)Q_\perp+fQ_z(f\sin\Theta+Q_\parallel)]    \\
        \frac{1}{2}(\hat\omega^2Q_\parallel-f^2Q_z\cos\Theta)       &   \frac{1}{2}[(\hat\omega^2-{N_0}^2)Q_\perp+fQ_z(f\sin\Theta+Q_\parallel)]    &   i\hat\omega(\hat\omega^2-{N_0}^2)
    \end{bmatrix}
\end{equation}
is the symmetric part of $\vec{\mathcal{M}}^{-1}$.
To make all first-order terms vanish, one would need
\begin{equation}
    \label{eq:hyp_a}
    2\grad a\cdot\vec{\mathcal{N}} = a\vec B_0.
\end{equation}
Supposing that this is possible, this would lead to
\begin{equation}
    \label{eq:normal}
    \grad\grad a\simeq \frac{a}{4{c_{\rm s}}^4}\left[\vec g_0\cdot(\vec{\mathcal{M}}^{-1})^T\vec{\mathcal{N}}^{-1}-2\beta\vec g_0\right]\otimes\left[\vec g_0\cdot(\vec{\mathcal{M}}^{-1})^T\vec{\mathcal{N}}^{-1}\right],
\end{equation}
where $\otimes$ denotes the outer product of two vectors, which is a tensor defined by $(\vec a\otimes \vec b)_{ij} = a_ib_j$.
The right-hand side of Eq.~\eqref{eq:normal} is not symmetric, and thus cannot be the double gradient of a function.
This proves that no function $a$ verifies Eq.~\eqref{eq:hyp_a}.
Instead we choose $a$ such that
\begin{equation}
    2\grad a\cdot\vec{\mathcal{N}} = a\vec B_{\rm s},
\end{equation}
where
\begin{equation}
    \vec B_{\rm s} = \frac{\vec g_0}{{c_{\rm s}}^2}\cdot\vec{\mathcal{N}}.
\end{equation}
This yields
\begin{equation}
    \grad a = a\frac{\vec g_0}{2{c_{\rm s}}^2},
\end{equation}
and Eq.~\eqref{eq:apsi} becomes
\begin{equation}
    \label{eq:wave_mod}
    -\vec{\mathcal{N}}:\grad\grad\psi + \vec B_{\rm a}\cdot\grad\psi + {k_{\rm c}}^2(\vec e_\parallel\cdot\vec{\mathcal{N}}\cdot\vec e_\parallel)\psi = 0,
\end{equation}
where
\begin{equation}
    \vec B_{\rm a}=\vec B_0-\vec B_{\rm s}=\frac{g_0}{{c_{\rm s}}^2\Gamma}\left[\hat\omega^2\left(f\sin\Theta+\frac{Q_\parallel}{2}\right)+\frac{f^2Q_z\cos\Theta}{2}\right]\vec e_\varphi
\end{equation}
and
\begin{equation}
    \label{eq:kc}
    {k_{\rm c}}^2=\frac{(1-2\beta){g_0}^2}{4{c_{\rm s}}^4}=\frac{(3+2\alpha-2\Gamma_1){g_0}^2}{4{c_{\rm s}}^4}.
\end{equation}
We note that the term in $\vec B_{\rm a}$ in Eq.~\eqref{eq:wave_mod} will be neglected when applying the JWKB approximation.

The expression for ${k_{\rm c}}^2$ given in Eq.~\eqref{eq:kc} is negative when $\alpha=0$ and $\Gamma_1>3/2$, which is the case in convective envelopes.
However, stars with a convective envelope always have a thin stably stratified surface layer.
In the case of a radiative envelope where $N_0$ becomes very large near the surface and scales as $1/{c_{\rm s}}^2$, computing the constant term is very lengthy, so we assume here that Eq.~(19) of Paper~I is still valid in the presence of differential rotation, namely
\begin{equation}
    \label{eq:kc_fin}
    {k_{\rm c}}^2=\frac{[(1+\alpha)^2-\beta^2]{g_0}^2}{4{c_{\rm s}}^4}.
\end{equation}

\section{Number of solutions of the equation $\Gamma=0$}
\label{sec:gamma}

In the case of uniform rotation, $\Gamma=0$ reduces to
\begin{equation}
    \omega^4-\omega^2({N_0}^2+f^2)+{N_0}^2f^2\cos^2\Theta=0,
\end{equation}
which always have two real solutions for $\omega^2$.

In presence of differential rotation, some calculations are required.
First, the discriminant of $\Gamma=0$ seen as a quadratic equation for $\omega^2$ reads
\begin{equation}
    \Delta = \left[{N_0}^2+f(f+Q_s)\right]^2+4f\cos\Theta\left[fQ_z(f\sin\Theta+Q_\parallel)-{N_0}^2(f\cos\Theta+Q_\perp)\right].
\end{equation}
It can be rewritten in terms of $Q_\parallel$ and $Q_\perp$:
\begin{equation}
    \begin{aligned}
    \Delta  &= {N_0}^4 + 2{N_0}^2\left[f^2(1-2\cos^2\Theta)+f(\sin\Theta Q_\parallel-\cos\Theta Q_\perp)\right] + f^4 + 2f^3\left[\sin\Theta Q_\parallel(1+2\cos^2\Theta)+\cos\Theta Q_\perp(1-2\sin^2\Theta)\right] \\
            &\quad + f^2\left[{Q_\parallel}^2(\sin^2\Theta+4\cos^2\Theta)-2\sin\Theta\cos\Theta Q_\parallel Q_\perp + \cos^2\Theta {Q_\perp}^2\right].
        \end{aligned}
\end{equation}
To go further, we need to consider the reduced discrimant of the equation $\Delta=0$ seen as a quadratic equation for ${N_0}^2$:
\begin{equation}
    \begin{aligned}
        \Delta' &= \left[f^2(1-2\cos^2\Theta)+f(\sin\Theta Q_\parallel-\cos\Theta Q_\perp)\right]^2 - f^4 - 2f^3\left[\sin\Theta Q_\parallel(1+2\cos^2\Theta)+\cos\Theta Q_\perp(1-2\sin^2\Theta)\right]    \\
                &\quad - f^2\left[{Q_\parallel}^2(\sin^2\Theta+4\cos^2\Theta)-2\sin\Theta\cos\Theta Q_\parallel Q_\perp + \cos^2\Theta {Q_\perp}^2\right].
    \end{aligned}
\end{equation}
After simplification, one finally obtains
\begin{equation}
    \Delta'=-4f^2\cos^2\Theta(f\sin\Theta+Q_\parallel)^2,
\end{equation}
which is always negative.
As a consequence, the discriminant $\Delta$ never changes sign, and remains always positive.
Thus, the quadratic equation $\Gamma=0$ always has two real solutions.

\section{Ray dynamics equations in spherical coordinates}
\label{sec:spherical}

The first step is to express the general eikonal equation~\eqref{eq:eik_axi} for axisymmetric waves in spherical coordinates:
\begin{equation}
    \omega^2 = \frac{A{k_r}^2+2Bk_rk_\theta+C{k_\theta}^2+D{k_{\rm c}}^2}{{k_r}^2+{k_\theta}^2 + {k_{\rm c}}^2},
\end{equation}
where
\begin{align}
    A   &= f^2\cos^2\theta+f\lbrace-\cos\theta\sin(\theta-\Theta)\cos(\theta-\Theta)Q_r + [\cos\Theta\cos(\theta-\Theta)+\cos\theta\sin^2(\theta-\Theta)]Q_\theta\rbrace+{N_0}^2\sin^2(\theta-\Theta),   \\
    B   &= -f^2\sin\theta\cos\theta-f[\cos\theta\cos^2(\theta-\Theta)Q_r+\sin\theta\sin^2(\theta-\Theta)Q_\theta] + {N_0}^2\sin(\theta-\Theta)\cos(\theta-\Theta),  \\
    C   &= f^2\sin^2\theta+f\lbrace[\cos\Theta\sin(\theta-\Theta)+\sin\theta\cos^2(\theta-\Theta)]Q_r-\sin\theta\sin(\theta-\Theta)\cos(\theta-\Theta)Q_\theta\rbrace+{N_0}^2\cos^2(\theta-\Theta), \\
    D   &= f^2\cos^2\Theta + f\cos\Theta[\sin(\theta-\Theta)Q_r+\cos(\theta-\Theta)Q_\theta].
\end{align}
Equations~\eqref{eq:dyn_r}-\eqref{eq:dyn_ktheta} then give
\begin{align}
    \frac{{\rm d}r}{{\rm d}t}           &= \frac{(A-\omega^2)k_r+Bk_\theta}{\omega(k^2+{k_{\rm c}}^2)},  \\
    \frac{{\rm d}\theta}{{\rm d}t}      &= \frac{Bk_r+(C-\omega^2)k_\theta}{r\omega(k^2+{k_{\rm c}}^2)},    \\
    \frac{{\rm d}k_r}{{\rm d}t}         &= -\frac{r\partial_rA{k_r}^2+2(r\partial_r B-B)k_rk_\theta+[r\partial_rC-2(C-\omega^2)]{k_\theta}^2+r\partial_rD{k_{\rm c}}^2+(D-\omega^2)r\partial_r({k_{\rm c}}^2)}{2r\omega(k^2+{k_{\rm c}}^2)},    \\ 
    \frac{{\rm d}k_\theta}{{\rm d}t}    &= -\frac{\partial_\theta A{k_r}^2+2(\partial_\theta B+A-\omega^2)k_rk_\theta+(\partial_\theta C+2B){k_\theta}^2+\partial_\theta D{k_{\rm c}}^2+(D-\omega^2)\partial_\theta({k_{\rm c}}^2)}{2r\omega(k^2+{k_{\rm c}}^2)}. 
\end{align}

\end{document}